\RequirePackage{ifpdf}
\documentclass[hyper,12pt,letterpaper]{JHEP3}
\pdfoutput =1
\usepackage{graphicx}
\usepackage{amsmath,amssymb,multirow,array}

\numberwithin{equation}{section}

\newcommand{\ben}{\begin{eqnarray}\displaystyle}
\newcommand{\een}{\end{eqnarray}}
\newcommand{\be}{\begin{equation}}

\newcommand{\ee}{\end{equation}}
\newcommand{\lb}{\left (}
\newcommand{\rb}{\right )}
\newcommand{\ltb}{\left [}
\newcommand{\rtb}{\right ]}
\newcommand{\ra}{\rightarrow}

\newcommand{\nn}{\nonumber}

\newcommand{\pa}{\partial}

\newcommand{\bc}{\begin{center}}
\newcommand{\ec}{\end{center}}

\newcommand{\Rmnum}[1]{\expandafter\@slowromancap\romannumeral #1@}

\newcommand{\ps}{{\psi(\bf r)}}
\newcommand{\A}{{\bf A}}
\newcommand{\B}{{\bf B}}
\newcommand{\js}{{\bf j}_s({\bf r})}

\def\cg2{\cos (\pi V)}
\def\sg2{\sin (\pi V)}
\def\cb2{\cos (\delta/2)}
\def\sb2{\sin (\delta/2)}

\def\sg{r_0^2 {\rm sinh}^2\gamma }

\def\cg{r_0^2 {\rm cosh}^2\gamma }

\def\[{\left [}
\def\]{\right ]}
\def\({\left (}
\def\){\right )}

\usepackage{float}
\usepackage[font=small,labelfont=bf]{caption}
\usepackage{mathrsfs}
\usepackage[parfill]{parskip}

\title{Chern-Simons Superconductor} \author{{\bf Nabamita
    Banerjee}$^1$\footnote{tpnb@iacs.res.in}, {\bf Suvankar
    Dutta}$^2$\footnote{suvankar@iiserb.ac.in}, {\bf Dibakar
    Roychowdhury}$^2$\footnote{dibakarphys@gmail.com,
    dibakar@iiserb.ac.in}\\ 
  $^1$Indian Association for the Cultivation of Science, 2A, B Raja S
  C Mullick Road, Jadavpur, Kolkata\\ 
  $^2$ Dept. of Physics, Indian Institute of Science Education and
  Research Bhopal, Bhopal 462 023, India }

\abstract{We study the effect of a bulk Chern-Simons (CS) term on ($
  3+1 $) dimensional type II superconductor in the context of the
  AdS/CFT correspondence. We holographically compute the super-current
  and find that it is non-local in nature. It receives non trivial
  corrections due to presence of the CS term. Considering a large
  limit of a parameter $\lambda$ (we call this limit as long wave
  length limit), which is effectively the high temperature limit of
  the theory, we find that this non-local super-current boils down to
  a local quantity. The leading term (without the CS term) of this
  current matches with the result of Ginzburg-Landau (GL) theory. We
  compute the effect of the CS term on GL current and find that the
  effect is highly suppressed at large temperature ($~\frac{1}{T^4}$).
  Finally, free energy of the vortex configuration has been
  calculated.  The free energy also receives non trivial correction at
  the order of $1/\lambda^2$ in the long wave length approximation.  }

\begin{document}

\section{Introduction and summary}

The phenomena superconductivity is characterised by two different
effects : $(i)$ Vanishing of resistivity, $(ii)$ macroscopic
diamagnetism below a critical temperature $T_c$. When temperature is
above $T_c$, the material behaves like a normal metal. At $T=T_c$, the
sample undergoes a second order phase transition from normal phase to
superconducting phase.

A phenomenological model proposed by Ginzburg and Landau in mid
nineties explains this phenomena qualitatively. According to this
model, in the normal phase free energy of the system is invariant
under a $global$ $U(1)$ symmetry. However, this symmetry is
spontaneously broken below the critical temperature $T_c$ and vacuum
expectation value of a charged scalar field $\Psi$ (order parameter of
phase transition) becomes non-zero (condensation). This spontaneous
breakdown of a global $U(1)$ symmetry characterises a second order
phase transition from normal phase to a superconducting phase. We
shall briefly review the Ginzburg-Landau (GL) model in section
$\S$\ref{GL}.

The basic idea of holographic superconductor came after phenomenal
papers by S. Gubser \cite{Gubser:2005ih}-\cite{Gubser:2008px}. In his
work he argued that ``coupling the Abelian Higgs model to gravity with
a negative cosmological constant leads to black holes which
spontaneously break the global $U(1)$ gauge invariance via a charged
scalar condensate slightly outside their horizon. This suggests that
black holes can superconduct.'' Based on this observation, Hartnoll,
Herzog and Horowitz proposed a gravity dual of $2+1$ dimensional
superconductor \cite{Hartnoll:2008vx}. In this paper they showed that
below a critical temperature, a charged condensate is formed via a
second order phase transition and the (DC) conductivity becomes
infinite. They considered a complex scalar field $\psi$, minimally
coupled with a $U(1)$ gauge field in four dimensional $AdS$ black
brane geometry. Ignoring the back reaction of the scalar field and
gauge field on black brane geometry, it was observed that, non-trivial
coupling between the gauge field and scalar field destabilises the
$\psi=0$ solution at some critical temperature $T_c$ and below that
temperature, the black hole admits a nontrivial scalar hair
solution. The effect of back reaction has been considered in
\cite{Hartnoll:2008kx}.

The effect of an external magnetic field on the charged scalar has
first been studied holographically in \cite{Hartnoll:2008kx}. The
authors added a magnetic field passing perpendicularly through the
material and obtained the London equation and magnetic penetration
depth. They also showed that these holographic superconductors are of
Type II. From their analysis it became quite evident that in presence
of a large magnetic field applied at low temperature, the system
behaves like a normal metal. As we decrease the magnetic field below a
critical value, the material develops superconducting vortices.

The vortex solution has further been studied in
\cite{Maeda:2009vf}\footnote{See also
  \cite{Albash:2008eh},\cite{Albash:2009iq},\cite{Salvio:2012at},\cite{Domenech:2010nf},\cite{Montull:2012fy},\cite{Montull:2009fe}.}. The authors studied
the bulk equations of motion for scalar field near the phase
transition line and showed that the solutions are holographic
realisation of Abrikosov's vortex lattice. The vortex solution is
thermodynamically favoured below a critical magnetic
field. Holographically calculation of free energy and superconducting
current turn out to be nonlocal functions. But these expressions
reduce to the known form of GL theory at long wavelength.

In this paper we study the properties of a $3+1$ dimensional
superconductor. The bulk theory is given by a five dimensional black
brane geometry. We add a Chern-Simons term in the bulk Lagrangian. Our
main goal is to understand the effect of the Chern-Simons term on the
properties of a holographic type II superconductor\footnote{See
  \cite{Zayas:2011dw} also.}. The Chern-Simons terms are interesting
as they source anomalous effects in the boundary theory even in
absence of the charged scalar field. There are three types of anomaly
in a quantum field theory : (1) associated with breakdown of a
classical symmetry at quantum level, (2) associated with breakdown of
classical gauge symmetry at quantum level and lastly (3) associated
with breakdown of a genuine symmetry of a quantum theory in presence
of background fields. There are many references on anomaly
\cite{Bertlmann:1996xk},\cite{Harvey:2005it},\cite{Bastianelli:2006rx},
\cite{Bilal:2008qx},\cite{Gynther:2010ed} and a recent one is
\cite{Banerjee:2012cr}, where the three different kinds of anomaly
have been explained with examples.  In our case, the presence of a
bulk CS term actually signifies the breakdown of a global $U(1)$
symmetry of the dual field theory.  We are interested to find if the
CS term introduces some anomalous effect in superconducting current
and vortex solution.

Our observations are following:
\begin{enumerate}
\item
The Chern-Simons term has no $direct$ effect on holographic vortex lattice
solution. 
\item The Chern-Simons term modifies the superconducting current. There
  are two kinds of effects. Holographically computed superconducting
  current turns out to be non-local. But in low frequency or long
  wavelength limit, it becomes a local quantity
  \cite{Maeda:2009vf}. In the absence of CS term the current takes the
  familiar form (in long wavelength approximation) of GL current
\be
J_i \sim \epsilon_i^j \pa_j \sigma(x,y) 
\ee
where $\sigma(x,y)$ is $|\psi|^2$.  However, there are two different
contributions of CS term on this current. $(a)$ In long wavelength
limit it only modifies the overall coefficient of the GL
current. $(b)$ The non-trivial local contribution that comes to the
superconducting current due to the presence of the CS term is at the
order of $1/\lambda^2$ (with respect to the leading term),
\be J_i \sim \epsilon_i^j \pa_j \sigma(x,y)
+\frac{\kappa^2}{\lambda^2} \epsilon_i^j \pa_j (\Delta \sigma(x,y)) ,
\ee 
where, $\kappa$ is the coefficient of the CS coupling and $\lambda$ is
defined in (\ref{lambdadf}). Later we shall see that $1/\lambda^2$
correction is effectively high temperature correction to the current.
\end{enumerate}

The paper is organized as follows.  In section $\S$\ref{GL} we give a
brief summary of the usual GL theory for type II superconductors. The
details of the dual gravitational description for $ (3+1) $
dimensional superconductor has been provided in section
$\S$\ref{sec:holomodel}. In section $\S$\ref{sec:boundcurrent}, using
the AdS/CFT prescription \cite{Maldacena:1997re}-\cite{Witten:1998qj}
we compute the boundary current in the presence of the CS term. In
section $\S$\ref{sec:freeenergy}, we compute the free energy for the
vortex configuration.  Finally we conclude in section
$\S$\ref{conclusion}.

The long appendix consists of details of some formulae that appear in
the expression of boundary current.

\section{Ginzburg-Landau theory of superconductor: a quick
  review} 
\label{GL} 

In 1950 V. L. Ginzburg and L. Landau proposed a phenomenological
theory of superconductor \cite{GL}. In this model, they
introduced a complex scalar field $\psi(\bf r)$ as an order parameter
of the system. $|\psi(\bf r)|^2$ represents the local density of
superconducting electrons. Ginzburg-Landau (GL) model was a
generalisation of London's theory of superconductors, where,
superconducting electrons' density $n_s$ was constant in
space. Because of the phenomenological foundation the GL theory did
not get much attention before 1959, when Gor'kov \cite{gorkov}
showed that GL theory is derivable as a special limiting case of
microscopic theory proposed by Bardeen, Cooper and Schrieffer in 1957
\cite{Bardeen:1957mv}.

In this section we plan to discuss important aspects of GL theory,
which we need to understand the results of this paper. We shall not go
into details of the derivations, rather we concentrate on the main
results of GL theory. Interested readers can find the details in
\cite{tinkham}.

In order to describe spatial non-uniformity of the order parameter $\psi(\bf
r)$, Ginzburg and Landau had to go beyond the idea of constant order
parameter (mean-field approximation), which can qualitatively describe
the phenomena of phase transition. They introduced a non-local free
energy of the system written in powers of $\ps$ and $\nabla\ps$,
\begin{equation}
F(\psi, \A) = \int d^3r \ltb \alpha |\ps|^2 +\frac{\beta}2 |\ps|^4
+\frac{1}{2 m^*} \bigg| \lb \frac1{i} \nabla -q \A \rb \ps\bigg|^2 +
\frac{B^2}{8\pi} \rtb .
\end{equation}
Thermodynamical equilibrium is attained minimising the free
energy. Minimising $F$ with respect to $\ps$ we get,
\begin{equation}
  \frac{1}{2m^*} \lb \frac1{i} \nabla -q \A \rb^2 \ps+\alpha \ \ps +
  \beta\ |\ps|^2 \ps =0 .
\end{equation}
However, minimisation with respect to the gauge field $\A$ gives rise
to the following equation, 
\begin{equation} -i \frac{q}{2m^*} \lb \ps
\nabla\psi^*(\bf r) - \psi^*(\bf r) \nabla \ps\rb + \frac{q^2}{m^*}
|\ps|^2 \A +\frac1{4\pi} \nabla \times {\bf B} =0 . 
 \end{equation} 
 Using Amp\'ere's law one can find a local super current ${\bf
   j}_s(\bf r)$ given by,
 \begin{eqnarray} \label{supcurrent}
\js &=& \frac1{4\pi} \nabla\times \B \nonumber \\
&=& i \frac{q}{2m^*}
  \lb \ps \nabla\psi^*(\bf r) - \psi^*(\bf r) \nabla \ps\rb -
  \frac{q^2}{m^*} |\ps|^2 \A .
\end{eqnarray}
As a consistency check, we find that in the limit of constant order
parameter ($\ps$) we obtain 
\begin{equation}
  \js = - \frac{q^2}{m^*}  |\ps|^2 \A \ .
\end{equation}
In this limit we expect that this supercurrent should reproduce the
London equation
\begin{equation}
  \js = - \frac{e^2 n_s}{m} |\ps|^2 \A 
\end{equation}
which requires,
\begin{equation}
|\ps|^2 = \frac{e^2 m^*}{q^2 m} n_s . 
\end{equation}

\begin{figure}[h]
	\centering
	\includegraphics[width=0.8\textwidth]{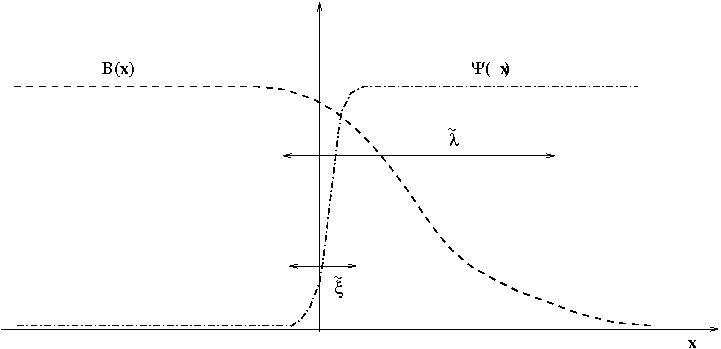}
        \caption{Behaviour of magnetic field and order parameter
          inside a type II syper-conductor.}\label{fig:II}
\end{figure}

We introduce two intrinsic length scales of this theory:
\begin{itemize}
\item
$\tilde \lambda$ : determines the penetration strength of external magnetic
field.
\item
$\tilde \xi$: Ginzburg-Landau coherence length : measures the length scale of
variation of the order parameter in the surface area and deep inside
the superconductor.
\end{itemize}
In the mean field approximation one can show that both the oder
parameters scale as
$$
\tilde\lambda, \tilde\xi \sim \frac{1}{\sqrt{T_c -T}}
$$
close to $T_c$. Therefore, we define a dimensionless quantity 
\be
\tilde \kappa = \frac{\tilde\lambda(T)}{\tilde\xi(T)}
\ee
which is a function of temperature. $\tilde \kappa$ is called Ginzburg-Landau
parameter. Depending on the value of $\tilde \kappa$ we classify two
different kinds of superconductors : (i) {\bf Type I} : $\tilde \kappa <
1/\sqrt{2}$ and (ii) {\bf Type II} : $\tilde \kappa> 1/\sqrt{2}$ (see
figure \ref{fig:II}). 

In type I superconductor the superconducting state disappears if the
applied magnetic field is greater than a critical value $H_c$. The
equilibrium state in the bulk of the superconductor is uniform. The
correct thermodynamic potential in presence of a magnetic field is
given by the Gibbs's free energy density ($\bf B$ is magnetic
induction inside the material),
\begin{eqnarray}
G/V &=& F/V - \frac1V \int d^3r \frac{{\bf H}\cdot {\bf B}}{4\pi} .
\end{eqnarray}
In the superconducting phase the magnetic field inside the
superconductor is zero, hence Gibb's potential is given by
\begin{eqnarray}
G_S/V &=& \alpha |\psi^2| +\frac{\beta}2 |\psi|^4 = -
\frac{\alpha^2}{2\beta},
\end{eqnarray}
Whereas, in the normal phase, $\psi$ vanishes, and magnetic field
penetrates the material. Hence, free energy for the normal state
is given by,
\begin{equation}
G_N/V = F_N/V -\frac{H \ B}{4\pi} = - \frac{H^2}{4\pi} .
\end{equation}
Therefore, there is a competition between these two phases and at
$H=H_c$ these two free energies become equal
\be H_c = \sqrt{4\pi
  \frac{\alpha^2}{\beta}} \ .
\ee
If the applied magnetic field $H<H_c$, superconducting state dominates
the thermodynamics and for $H>H_c$, normal state takes over. 

If the sample is very large (for example consider a thin slab) and the
applied magnetic field passes perpendicularly through the sample then
even for $H<H_c$ the flux penetrates through the sample (since going
around the large sample would cost much energy) forming small normal
strips. Therefore, in type I super conductor one can see such
``intermediate states''.

\subsection{Type II superconductor: vortex solution}

Type II superconductor is characterised by the formation of magnetic
vortices in an applied magnetic field. Vortices form in the
superconducting material when $\tilde \kappa > 1/\sqrt{2}$.

Consider an area $S$ with boundary $\partial S$. We assume that
at least the boundary $\partial S$ lies inside the
superconductor. Therefore, net flux passing through this area is given
by,
\be
\Phi = \oint_{\partial S} d{\bf s}\cdot {\bf A} .
\ee
Using equation (\ref{supcurrent}) one can write,
\be
\Phi = -\frac{m}{e^2 n_s} \oint_{\partial S} d{\bf s}\cdot\js
-\frac{1}{2e}\oint_{\partial S} d{\bf s}\cdot{\bf \nabla} \phi 
\ee
where, $\phi$ is the phase of the complex field $\psi$. The last term
is an integer multiple of $2\pi$,
\be
\oint_{\partial S} d{\bf s}.{\bf \nabla} \phi = - 2\pi n, \qquad n \in
\mathbb{Z} .
\ee
Thus we see that
\be
\Phi' = \Phi +\frac{m}{e^2 n_s} \oint_{\partial S} d{\bf s}\cdot\js  =
n \Phi_0, \qquad \Phi_0 = {\pi\over e} 
\ee
is quantised. The quantity $\Phi'$ is called fluxoid. Therefore, the
magnetic fields penetrate the superconductor in quantised way, $i.e.$
the magnetic fluxoid is integral multiple of minimum fluxoid
$\Phi_0$. The penetration takes place in terms of vortices or vortex
lines. We consider the direction of a vortex is along the direction of
the magnetic field. As we go around a vortex anticlockwise the phase
of $\ps$ changes by an amount $-2\pi$. At the centre of a vortex the
phase is not defined. Therefor for a consistent solution for $\ps$ we
consider $\ps=0$ at the core. In cylindrical coordinate system $(\rho,
\varphi,z)$ (vortex is along $z$ direction), we find that,
\ben\label{bctypeII}
\psi(\rho=0) =0\nn \\
|\psi(\rho\ra \infty)| =\psi_0\nn\\
\B(\rho\ra \infty) =0.
\een
Since the system has cylindrical symmetry, we can choose,
\ben
\ps &=& \psi_{arg}(\rho) e^{-i \varphi}\nn\\
\js &=& \hat{\varphi} j_s(\rho)\nn\\
\B({\bf r}) &=& \hat{z} B(\rho) 
\een
Choosing the vector potential along $\hat{\varphi}$ direction one can solve
the Landau-Ginzburg equation in cylindrical coordinate along with the
expression for supercurrent (Ampere's law) under the boundary
conditions given in (\ref{bctypeII}). The equations are highly
non-linear. A numerical solution gives the following plot for the
magnetic field and condensation. 


In type-II superconductor, nothing interesting happens unless the
applied magnetic field is greater that the lower critical magnetic
field $H_{c1}$. For $H<H_{c1}$ the sample exhibits Meisner
phase. The first vortex enters the superconductor at $H=H_{c1}$. For a
type II superconductor with $\tilde \kappa>>1$ lower critical magnetic field
is given by,
\be
H_{c1} = \frac{H_c}{\sqrt{2}} \frac{\ln \tilde \kappa}{\tilde \kappa} .
\ee

\subsubsection{Vortex lattice}

The flux in superconductor should enter as periodic triangular lattice
to minimise the free energy of vortex state. However, the vortex state
disappear as we increase the external magnetic field beyond the upper
critical value $H_{c2}$. For $H>H_{c2}$, the sample becomes normal
metal. 

Based on Ginzburg-Landau theory Abrikosov proposed a solution for this
vortex lattice. Abrikosov’s results are quantitatively valid only
near $H_{c2}$. In his construction he assumed that the magnetic flux
density is uniform, which is valid only near $H_{c2}$.

For a constant $\B$ along $\hat z$ direction, we write $A_y= H x$. The
linearised Ginzburg-Landau equation becomes
\ben \label{LGeqn_vortex}
\frac{1}{2m^*}\lb  {1\over i} \nabla + {2 e H}\hat y x \rb^2\psi +
\alpha \ \psi =0 \nn\\
\ltb -\nabla^2 -\frac{4\pi i}{\Phi_0}H x {\partial \over \partial y} +
\lb \frac{2\pi H}{\Phi_0} \rb^2 x^2 \rtb \psi = \frac1{\xi^2} \psi
\een
where, 
\be
\xi^2(T) = - \frac{1}{2 m^* \alpha(T)}
\ee
This equation looks like Schr\"odinger equation with nontrivial
potential along $x$ direction. The particle behaves like a free
particle in $y$ and $z$ direction. Therefore we consider the following
ansatz for $\psi$
\be
\psi = e^{i k_y y} e^{i k_z z} f(x) .
\ee
Therefore equation (\ref{LGeqn_vortex}) becomes,
\be\label{vortexeqn}
-f''(x) + \lb \frac{2\pi H}{\Phi_0}\rb^2 (x-x_0)^2 f(x) =\lb {1\over
  \xi^2} - k_z^2 \rb f(x) 
\ee
where,
\be
x_0 = \frac{k_y \Phi_0}{2 \pi H} .
\ee
This equation is exactly same as one dimensional quantum harmonic
oscillator with frequency $\omega_c = 2H e/m^* c$ and minimum of the
potential shifted by an amount $x_0$. This is an eigenvalue equation
and the solution exists if 
\be
H = \frac{\Phi_0}{2\pi (2n+1)}\lb {\frac{1}{\xi^2}-k_{z}^{2}}\rb .
\ee
This equation says that there exists an upper limit of the magnetic
field which corresponds to $n=0$ and $k_z=0$. This is the upper
critical magnetic field $H_{c2}$. Therefore, we find that,
\be
H_{c2} = \frac{\Phi_0}{2\pi \xi^2(T)}. 
\ee
One can also write the relation between upper critical magnetic field
and the thermodynamic magnetic field,
\be
H_{c2} = \sqrt2 \kappa H_c . 
\ee

We consider the lowest energy solution ($k_z=0, \ n=0$),
\be
\psi_0(x,y) = e^{i k_y y} f_0(x)\label{psinot}.
\ee
Abrikosov assumed that the solution for $H\lesssim H_{c2}$ is periodic
in $x$ and $y$ direction. $y$ direction has a period $a_y$, hence,
\be
k_y = \frac{2\pi}{a_y} l, \quad l \in \mathbb{Z} .
\ee
Therefore, the harmonic oscillator is centered at
\be
x_0 = \frac{\Phi_0}{H a_y}l
\ee
and the solution is given by,
\be 
\psi_0(x,y) = {\cal N} \exp{\lb i\frac{2\pi y}{a_y}l \rb}\exp{\lb -\frac12
  m^* \omega_c \ltb x-\frac{\Phi_0}{H a_y}l \rtb^2\rb} .
\ee
In the vicinity of $H_{c2}$, we can write
\be
\omega_c m^* = \frac{2 H e}{c}\approx \frac{2H_{c2}e}{c} =
\frac{1}{\xi(T)^2} .
\ee
Thus,
\be
\psi_0(x,y) = {\cal N} \exp{\lb i\frac{2\pi y}{a_y}l \rb}\exp{\lb
  -\frac1{2\xi(T)^2}  \ltb x-\frac{\Phi_0}{H_{c2} a_y}l \rtb^2\rb} .
\ee
The most general solution, therefore, is given by linear superposition
for different $l$
\be
\psi_0(x,y) = \sum_{l=-\infty}^{\infty} c_l \exp{\lb i\frac{2\pi y}{a_y}l \rb}\exp{\lb
  -\frac1{2\xi(T)^2}  \ltb x-\frac{2\pi l}{a_y}\xi(T)^2 \rtb^2\rb} .
\ee
We can write this solution in terms of $elliptic \ theta \ function$,
\be
\psi_0(x,y) = e^{-\frac{x^2}{2\xi^2}} \vartheta_3(v,\tau)\label{psi0} .
\ee
The theta function is defined by,
\be
\vartheta_3 = \sum_{l=-\infty}^{\infty} q^{l^2} z^{2 l}, \quad q=
e^{\tau \pi i}, \ \ z= e^{i\pi v} 
\ee
where,
\be
v= \frac{-i x +y}{a_y}, \qquad \tau = \frac{2\pi i - a_x}{a_y}
\xi(T)^2. 
\ee
Here we have written $c_l$ in terms of new constant $a_x$,
\be
c_l = \exp\lb - \frac{i \pi a_x}{a_y^2} \xi(T)^2 l^2\rb
\ee

From the periodicity pf theta function,
\ben
\vartheta_3(v+1,\tau) &=& \vartheta_3(v,\tau)\nn \\
 \vartheta_3(v+\tau,\tau) &=& e^{-2\pi i (v+\tau/2)}
 \vartheta_3(v,\tau) 
\een
one can check that,
\ben
\psi_0(x,y+a_y) &=& \psi_0(x,y)\nn \\
\psi_0(x+\frac{2\pi \xi^2}{a_y}, y +\frac{a_x \xi^2}{a_y}) &=&\exp\ltb
2\pi i \lb \frac{y}{a_y} + \frac{a_x \xi^2}{2 a_y^2}\rb \rtb
\psi_0(x,y) .
\een

Thus $\sigma(x,y)= |\psi_0|^2$ represents a lattice in which the
fundamental region is spanned by two vectors ${\bf b_1} =
a_y \partial_y$ and ${\bf b_2} = \frac{2\pi \xi^2}{a_y} \partial_x +
\frac{a_x \xi^2}{a_y} \partial_y$. The area of the fundamental region
is given by $2 \pi \xi^2$. Therefore the total flus passion through
the fundamental region is given by,
 \be
 xH_{c2} \times Area = 2\pi \xi^2 H_{c2} =\Phi_0 .
\ee

The current evaluated for this vortex solution is given by,
\be
j_i = \epsilon^j_i \pa_j \sigma(x,y) .
\ee

\section{Holographic model for $3+1$ dimensional
  superconductor} \label{sec:holomodel} 

We consider the following five dimensional action with a Chern-Simons term
\begin{eqnarray}
  S = \frac{1}{2l_p^{3}}\int d^{5}x\sqrt{-g}\left( 12 + R -
    \frac{1}{4}F^{2}+\frac{\kappa}{3}\epsilon^{abcde}A_aF_{bc}F_{de}-|\nabla_a
    \psi -iA_a\psi |^{2} - m^{2}|\psi |^{2} \right) \label{action}
\end{eqnarray}

We consider gauge field and scalar field in the probe approximation
and treat the effect of Chern-Simons term perturbatively ($\kappa$
small). The background (asymptotically) $AdS$ black brane solution is
given by\footnote{ Our choice of coordinate is special, so that $u
  f(u)$ is finite at the boundary $u \rightarrow 0$. With this choice,
  the boundary current (\ref{bound-current}) is finite. This choice
    differs from the usual AdS metric on Poincare patch (which we
    usually use to define the dimensions of various fields) along the
    radial direction as $u=r_{pc}^2$.},
\be
ds^2 = -f(u) dt^2 + \frac{r_0^2}{4u^3 f(u)} du^2 + \frac{r_0^2}{u}
(dx^2 +dy^2+dz^2), \qquad f(u) = \frac{r_0^2}u (1-u^2) . 
\ee
The horizon is located at $u=1$ and asymptotic boundary is at $u\ra
0$. 
The temperature of this black brane is given by,
\be
T = \frac{r_0}{\pi}.
\ee

We consider the following ansatz for the gauge field and scalar field,
\ben
A &=& {(\phi(u), 0,0, A_y(x), P(u)})\nn\\
\psi &=& \psi(u,x,y). 
\een

Using this ansatz, the equation of motion for the gauge field turns out to be,
\ben \label{gaugeeom}
&&\phi ''(u)-\frac{r_0^2 \phi (u) \psi (u,x,y)^2}{2 u^3
  f(u)}+\frac{12 \kappa {A_y}'(x)
  P'(u)}{r_0^2} = 0 \nn\\
&&u {A_y}''(x)-2 r_0^2 {A_y}(x) \psi (u,x,y)^2 = 0\nn\\
&&P''(u) +\frac{P'(u)}{u}+\frac{f'(u) P'(u)}{f(u)} -\frac{r_0^2 P(u)
  \psi (u,x,y)^2}{2 u^3 f(u)}+\frac{12 \kappa {A_y}'(x) \phi '(u)}{u
  f(u)} = 0. 
\een
Likewise, the equation for scalar field is given by,
\ben\label{scalareom}
\lb \frac{\partial^2}{\pa u^2} + \frac{f'(u)}{f(u)} \frac{\pa}{\pa u} +
\frac{r_0^2 \phi (u)^2-f(u) \left(m^2 r_0^2+u P(u)^2\right)}{4 u^3
  f(u)^2} \rb \psi &&\nn\\
 +\frac1{4u^2 f(u)} \lb \Delta -2 i
A_y(x)-A_y(x)^2\rb \psi &=& 0 .
\een

Our goal is to find a consistent solution to these equations. We
impose the following boundary conditions in order to solve these
equations. 
\begin{itemize}
\item We consider the background temperature stable. We neglect the
  effect of the scalar field and the gauge field on the background. Therefore, the
  black brane temperature is kept fixed in our calculation.
\item
Asymptotic value of the time component of the gauge field $\mu =
A_t(0)$, which behaves like a chemical potential of the boundary field
theory is kept fixed in our study. This means we are considering the
boundary field theory in grand canonical ensemble. 
\item
We also demand that the norm of the gauge field is finite at the
horizon. This impose a second boundary condition on $A_t(u): A_t(u=1)
=0$. 
\item
Non-normalisable mode of the gauge field $i.e.$ $xy$ component of the
field strength deforms the boundary. The asymptotic value of $F_{xy}$
gives the external magnetic field $H$.  
\item
The scalar field has different fall off at $AdS$ boundary for
different values of $m^2$. We consider the mass of the scalar field to
satisfy $BF$ bound in five dimensions $i.e.$ $m^2 =-3$. In this case
the scalar field goes as,
\be
\psi \sim \alpha_1 \sqrt{u} + \alpha_2 u^{3/2} \label{bcpsi}.
\ee
We see that both the modes are normalisable. Therefore the
coefficients $\alpha_1$ and $\alpha_2$ correspond to the expectation
value of some dual operators with dimensions $\Delta=1$ and $\Delta=3$
respectively. In our case we turn off $\alpha_1$ and
consider only the faster falling mode $\alpha_2$. 

\end{itemize}

The external magnetic field $H$ is the tuning parameter of our
system. We shall control the value of this magnetic field from out
side and study the effect on condensation and currents in the boundary
theory. We expect that the scalar field begins to condensate below a
critical magnetic field. We call this magnetic field $H_{c2}$ and
define a parameter $\varepsilon = \frac{H_{c2}-H}{H_{c2}}$ with
$\varepsilon \ll 1$. Our goal is to find solution to the above
equations of motion in power series of $\epsilon$. We expand fields in
the following way,
\ben
\psi(u,x,y) &=& \varepsilon^{1/2} \psi_1(u,x,y) + \varepsilon^{3/2}
\psi_2(u,x,y) + \cdots \nn\\
A_a (u,x,y) &=& A_{a}^{(0)} + \varepsilon A_{a}^{(1)}(u,x,y)+\cdots \label{pertb}.
\een

Solution in $\epsilon\ra 0$ limit is given by,
\ben
A_t^{(0)} &=&\phi(u)= \mu  (1-u)-\frac{144 \kappa ^2 \mu
  H_{\text{c2}}^2 (2u \log 2-(u+1) \log 
   (u+1))}{r_0^4} + {\cal O}(\kappa^4) ,\nn\\
A_y^{(0)}(x) &=& H_{c2} x , \nn\\
A_z^{(0)}(u) &=& P(u)=\kappa  \left(C_2-\frac{12 \mu  H_{\text{c2}} \log
    (u+1)}{r_0^2}\right)+{\cal O}(\kappa^3) \label{solgauge}. 
\een
The above solution is perturbative in the CS coefficient
$\kappa$. Since we are treating the gauge (and scalar field) as probe,
we begin with the AdS Schwarzschild solution as feed and solve for the
gauge fields perturbatively in $\kappa$ up to first non trivial order,
setting the scalar to zero. The $\kappa=0$ solution for $A_z$ is a
constant and we set that to zero\footnote{This is fine as we do get a
non zero chemical potential for $A_z$ at $\kappa$ order.}.

We substitute these leading order solutions to equation for the scalar
field (\ref{scalareom}) to solve $\psi_1$. We write the scalar field as a product of
two independent parts and since the $y$ direction is a
free direction we take the following ansatz for $\psi_1$,
$$
\psi_1 = \rho(u) X(x) e^{i p y} . 
$$
Substituting this in the scalar field equation we get
\ben\label{rhoeqn}
\frac{f'(u) \rho'(u)}{f(u)}-\frac{m^2 r_0^2 \rho(u)}{4 u^3
  f(u)}-\frac{P(u)^2 \rho(u)}{4 u^2    f(u)}+\frac{r_0^2 \rho(u) \phi
  (u)^2}{4 u^3 f(u)^2}+\rho''(u)&=&-\bar \lambda  H\frac{ \rho(u)}{4 u^2
  f(u)}\label{rad} \\
X''(x) -H^2\left(x-\frac{p}H\right)^2X(x)&=&
-\bar \lambda X(x),
\label{Xeom}
\een 
where, $\bar\lambda$ is a an arbitrary constant.  Comparing
equation (\ref{Xeom}) with equation (\ref{vortexeqn}) we find that in
our case $\Phi_0=2\pi$, $\bar \lambda = \frac{1}{\xi^2}$ ($k_z=0$, in
our case). The maximum magnetic field below which vortex solution
exists given by, $H_{c2}=\bar\lambda$. Equation (\ref{rhoeqn}) has a
nontrivial solution only when $\bar\lambda$ (or $H_{c2}$) is a
nontrivial function of temperature $T$. There exists a critical
temperature $T_c$ at which the solution $\rho=0$ becomes marginally
stable. Below that temperature, the scalar field condensates and we
get a non-trivial profile satisfying the corresponding boundary
conditions. Thus we see that, although the CS term does not have any
effect on vortex equation but it has non-trivial effect on
condensation of the scalar mode.

\section{Boundary current}
\label{sec:boundcurrent}

In this section, using the AdS/CFT prescription \cite{Gynther:2010ed},
we compute the boundary current ($ J^{a} $) for the 4-dimensional
chiral $ U(1) $ gauge theory living on the boundary of the
5-dimensional bulk AdS space time. In order to compute the current
coupled to the $ U(1) $ gauge field, we evaluate the \textit{onshell}
action to the first order in the gauge fluctuations and finally using
the AdS/CFT dictionary we obtain\footnote{One can show that this
  current is anomalous as it has a non zero divergence \cite{Gynther:2010ed}. Note
  that this is a global anomaly associated with the $ U(1) $
  current. It will be remarkable, if we can indeed manage to write
  this \textit{anomalous} current as a local function of the vortex
  solution as prescribed by the standard Ginzburg-Landau theory.},
\begin{eqnarray}
  J^a =\frac{\delta S}{\delta A_a}|_{u\rightarrow 0}=
  \frac{\sqrt{-g}}{16\pi G}\left( F^{a u}+\frac{4 \kappa}{3 \sqrt{-g}} 
    \epsilon^{uabcd}A_{b}F_{cd}\right)|_{u\rightarrow 0}.
\end{eqnarray}
Setting $ a = i= (x,y,z) $, this finally yields\footnote{According to
  the conventions of the present paper we have considered
  $\epsilon^{01234}=1$, in local lorentz frame.} \footnote{Note that
  since the boundary theory lives on a flat background with positive
  signature, therefore $ J^{i}=J_{i} $.},
\begin{eqnarray} \label{bound-current}
  J_{i}=\frac{\sqrt{-g}\varepsilon}{16\pi G}\left( F^{(1)}_{iu}g^{ii}g^{uu} +
    \frac{4\kappa}{3 \sqrt{-g}}\epsilon^{uibcd}\left(A_{b}^{(1)}F_{cd}^{(0)}+A_{b}^{(0)}F_{cd}^{(1)} \right)
  \right)|_{u\rightarrow 0} +\mathcal{O}(\varepsilon^{2})\label{j}.
\end{eqnarray}
Our next task is to explicitly compute (\ref{j}) for the first order
gauge fluctuations. This implies that we
 need to solve the Maxwell equations at first order which is given by,
\begin{eqnarray}
\nabla_{b}F^{(1)ab}-2\frac{\kappa}{\sqrt{-g}}
\epsilon^{abcde}F^{(0)}_{bc}F^{(1)}_{de}= j^{(1)a}\label{f}. 
\end{eqnarray}

In order to solve (\ref{f}), we express it explicitly for the temporal as well as 
spatial parts of the gauge field. Writing (\ref{f}) explicitly for the
temporal part we find,
\begin{eqnarray}
  L_t A^{(1)}_{t}-\frac{16\kappa
    u^2f(u)}{r_0^{2}}(\partial_{u}A^{(0)}_{z}F^{(1)}_{xy}+H_{c2} \partial_{u}A^{(1)}_{z})
  =\frac{2r_0^{2}}{u}A^{(0)}_{t}|\psi_{1}|^{2}\label{at}.
  \end{eqnarray}
  On the other hand, for the spatial components we enumerate the equations
  below\footnote{In order to arrive at
  these equations we choose the gauge $ A_u =0 $. Also we have
  exploited the residual gauge symmetry $ A^{(1)}_i\rightarrow
  A^{(1)}_i-\partial_i\Lambda(x,y) $, in order to fix the gauge
  $ \partial_{x}A^{(1)}_{x}+\partial_{y}A^{(1)}_{y}=0 $.}
  \begin{eqnarray}
  L_s A^{(1)}_{x}-16 \kappa u
  (\partial_{u}A^{(0)}_{t}\partial_{y}A^{(1)}_{z}- \partial_u
  A^{(0)}_{z}\partial_{y}A^{(1)}_{t})
  &=&-\frac{r_0^{4}}{u^{2}}j^{(1)x}\nonumber \\ 
  L_s A^{(1)}_{y}-16 \kappa u(\partial_u
  A^{(0)}_{z}\partial_{x}A^{(1)}_{t}-\partial_{u}A^{(0)}_{t}\partial_{x}A^{(1)}_{z})
  &=&-\frac{r_0^{4}}{ u^{2}}j^{(1)y}\nonumber \\
  L_s A^{(1)}_{z}-16 \kappa u
  (\partial_{u}A^{(0)}_{t}F^{(1)}_{xy}+H_{c2} \partial_{u}A^{(1)}_{t})
  &=&-\frac{r_0^{4}}{ u^{2}}j^{(1)z}\label{l}.
\end{eqnarray}
Where we have defined two following linear operators as,
\begin{eqnarray}
L_t&=&(4 f(u)u^{2}\partial_{u}^{2}+\Delta), \quad L_s=  \left( 4
  u \partial_{u}(uf(u)\partial_u )+\Delta \right)\label{ls}. 
\end{eqnarray}
Note that in the above we have denoted the first order fluctuations in the time
as well as as in the spatial components of the gauge field as $ A^{(1)}_{t}  $ 
and  $ A^{(1)}_{i}  $ ($ i= x,y,z $) respectively. Remember these fluctuations
are arising solely due to the presence of a nontrivial profile of scalar hair,
which is basically a perturbation that one can identify as the \textit{order parameter}. 

Since our bulk theory contains a non trivial coupling between the
scalar field and the $ U(1) $ gauge sector, therefore the reader might
guess that turning on fluctuations both in the scalar as well as
in the gauge sector eventually make it quite
difficult to solve these equations exactly. This fact is indeed quite
evident from the above set of equations (\ref{at}) and (\ref{l}). Therefore in our
computations we aim to solve these equations pertubatively in $ \kappa
$ and we keep terms upto quadratic order in $ \kappa $\footnote{The reason for keeping terms upto quadratic order
in $ \kappa $ follows from the fact that in the previous section we have already retained
our solutions for the gauge fields upto quadratic order in the parameter $ \kappa $ (see (\ref{solgauge})). }.
In order to solve (\ref{at}) and (\ref{l}) perturbatively in $ \kappa $, we consider
the following expansion of the gauge field
\begin{equation}
  A^{(m)}_{n}= \mathcal{A}^{(m)(\kappa^{(0)})}_{n}+\kappa
  \mathcal{A}^{(m)(\kappa^{(1)})}_{n}+\kappa^{2}
  \mathcal{A}^{(m)(\kappa^{(2)})}_{n}+\mathcal{O}(\kappa^{3})\label{a}. 
\end{equation}
Also from the structure of the radial equation (\ref{rad}), it is indeed 
evident that we may express the solution perturbatively
in $ \kappa $ as,
\begin{eqnarray}
\rho_{0}=\rho_{0}^{(\kappa^{(0)})}+\kappa^{2}
\rho_{0}^{(\kappa^{(2)})}+\mathcal{O}(\kappa^{4})\label{rho}.
\end{eqnarray}
Note that we have used two different indices in the superscript, both of which correspond 
to different order of perturbations. Index ($ m $)
corresponds to fluctuations at different order due to the presence of a non trivial value 
of the order parameter. This is basically the perturbation in $\varepsilon$ that we have
already mentioned earlier in (\ref{pertb}). On the other hand, the superscript ($ \kappa^{(n)} $)
stands for different terms in the perturbative expansion in $ \kappa $.

We now substitute (\ref{a}) and (\ref{rho}) into  (\ref{at}) and (\ref{l}) in order to
solve these equations order by order in $ \kappa $. Let us first consider the
zeroth order equation.

\textbf{Equations in the zeroth order in $ \kappa $:}

We first write down (\ref{at}) and (\ref{l}) for the zeroth order, which take the following form,
\begin{eqnarray}
  L_t\mathcal{A}_{t}^{(1)(\kappa^{(0)})}&=&\frac{2r_0^{2}\rho_{0}^{2(\kappa^{(0)})}\sigma ({\bf{x}})}{u}\mathcal{A}_{t}^{(0)(\kappa^{(0)})}\nonumber\\
  L_s\mathcal{A}^{(1)(\kappa^{(0)})}_{x}&=& \frac{r_0^{2}\rho_{0}^{2(\kappa^{(0)})}}{u}\epsilon_{x}^{y}\partial_{y}\sigma(\textbf{x})\nonumber\\
  L_s\mathcal{A}^{(1)(\kappa^{(0)})}_{y}&=& \frac{r_0^{2}\rho_{0}^{2(\kappa^{(0)})}}{u}\epsilon_{y}^{x}\partial_{x}\sigma(\textbf{x})\nonumber\\
  L_s\mathcal{A}^{(1)(\kappa^{(0)})}_{z}&=&0\label{l1}
\end{eqnarray}
where we have used the fact that $ \epsilon_{ij} $ is anti symmetric,
satisfying $ \epsilon_{12}=-\epsilon_{21}=1 $.  Note that on the
r.h.s. of the above set of equations (\ref{l1}), we have
expressed the source $ j_{i}^{(1)} $ (associated with the first
order fluctuations of the non zero condensate) in terms of  $ \sigma(\textbf{x})(=|\psi_0|^{2}) $
that corresponds to the vortex solution (corresponding to a \textit{triangular} lattice)
in the ($ x,y $) plane. 

Since these are inhomogeneous differential equations with a source
term on the r.h.s. of it, therefore the solutions of (\ref{l1}) could
be written in terms of Green's functions, that satisfy the appropriate
boundary conditions at the AdS boundary.  The solutions at the zeroth
order are given below\footnote{Note that at the zeroth order in $
  \kappa $ the $ z $ component of the gauge field does not posses any
  solution, which is consistent with the findings of the previous
  section.}.
\begin{eqnarray}
\mathcal{A}_{t}^{(1)(\kappa^{(0)})}&=&-(2r_0^{2})\int_{0}^{1}du^{'}
\frac{\rho_{0}^{2(\kappa^{(0)})}(u^{'})}{u'}\mathcal{A}_{t}^{(0)(\kappa^{(0)})}(u^{'})\times \int d\textbf{x}^{'}\mathcal{G}_t(u,u^{'};\textbf{x},\textbf{x}')\sigma(\textbf{x}')\nonumber\\
\mathcal{A}_{i}^{(1)(\kappa^{(0)})}&=&a_i(\textbf{x})-r_0^{2}\epsilon_{i}^{j}
\int_{0}^{1}du^{'}\frac{\rho_{0}^{2(\kappa^{(0)})}(u^{'})}{u'}\times \int d\textbf{x}^{'}\mathcal{G}_{s}(u,u^{'};\textbf{x},\textbf{x}')\partial_{j}\sigma(\textbf{x}')\nonumber\\
\mathcal{A}^{(1)(\kappa^{(0)})}_{z}&=&0\label{zeroth}.
\end{eqnarray}
Note that $ a_i(\textbf{x}) $ written above, corresponds to a
homogeneous solution of (\ref{l}) which is solely responsible for
giving rise to the critical magnetic field ($ H_{c2}=
\epsilon_{ij}\partial_ia_j$) at the boundary of the AdS. $ \mathcal{G}_t(u,u^{'};\textbf{x},\textbf{x}') $
and $ \mathcal{G}_{s}(u,u^{'};\textbf{x},\textbf{x}') $ are the 
Green's functions of the above set of equations (\ref{l1}) that satisfy
the following equations of motion
\begin{eqnarray}
L_t \mathcal{G}_t(u,u^{'};\textbf{x},\textbf{x}')&=&-\delta(u-u')\delta(\textbf{x}-\textbf{x}')\nonumber\\
L_s \mathcal{G}_s(u,u^{'};\textbf{x},\textbf{x}')&=&-\delta(u-u')\delta(\textbf{x}-\textbf{x}')
\end{eqnarray}
along with the following (Dirichlet) boundary conditions at the AdS boundary. 
\begin{eqnarray}
\mathcal{G}_t(u,u^{'};\textbf{x},\textbf{x}')|_{u=0}=\mathcal{G}_t(u,u^{'};\textbf{x},\textbf{x}^{'})|_{u=1}=0\nonumber\\
\mathcal{G}_{s}(u,u^{'};\textbf{x},\textbf{x}')|_{u=0}=uf(u)\partial_u\mathcal{G}_{s}(u,u^{'};\textbf{x},\textbf{x}^{'})|_{u=1}=0\label{bc1}.
\end{eqnarray}
These boundary conditions also ensure the following two things:

(1) It means that the source $ \mu(=A_t(u=0)) $ is kept fixed at the boundary.

 (2) We have a uniform magnetic field ($ H_{c2} $) at the boundary of the AdS.

Next we almost follow the same procedure in order to find solutions of
(\ref{l}) for the leading order as well as next to the leading order
in $ \kappa $. Let us first note down the equations in the leading
order in $ \kappa $. 

\textbf{Equations in the leading order in $ \kappa $:}
\begin{eqnarray}
L_t\mathcal{A}_{t}^{(1)(\kappa^{(1)})}-\frac{16  u^{2}f(u)}{r_0^{2}}(\partial_{u}\mathcal{A}^{(0)(\kappa^{(0)})}_{z}\mathcal{F}^{(1)(\kappa^{(0)})}_{xy}+H_{c2} \partial_{u}\mathcal{A}^{(1)(\kappa^{(0)})}_{z})&=&\frac{2r_0^{2}\rho_{0}^{2(\kappa^{(0)})}\sigma ({\bf{x}})}{u}\mathcal{A}_{t}^{(0)(\kappa^{(1)})}\nonumber\\
L_s\mathcal{A}^{(1)(\kappa^{(1)})}_{x}- 16  u(\partial_{u}\mathcal{A}^{(0)(\kappa^{(0)})}_{t}\partial_{y}\mathcal{A}^{(1)
(\kappa^{(0)})}_{z}-\partial_u \mathcal{A}^{(0)(\kappa^{(0)})}_{z}\partial_{y}\mathcal{A}^{(1)(\kappa^{(0)})}_{t})&=&0\nonumber\\
L_s\mathcal{A}^{(1)(\kappa^{(1)})}_{y}-16 u(\partial_u \mathcal{A}^{(0)(\kappa^{(0)})}_{z}\partial_{x}\mathcal{A}^{(1)(\kappa^{(0)})}_{t}-\partial_{u}\mathcal{A}^{(0)(\kappa^{(0)})}_{t}\partial_{x}\mathcal{A}^{(1)(\kappa^{(0)})}_{z})&=&0\nonumber\\
L_s\mathcal{A}^{(1)(\kappa^{(1)})}_{z}-16 u(\partial_{u}\mathcal{A}^{(0)(\kappa^{(0)})}_{t}\mathcal{F}^{(1)(\kappa^{(0)})}_{xy}+H_{c2} \partial_{u}\mathcal{A}^{(1)(\kappa^{(0)})}_{t})&=&\frac{2r_0^{2}\rho_{0}^{2(\kappa^{(0)})}\sigma ({\bf{x}})}{u}\mathcal{A}_{z}^{(0)(\kappa^{(1)})}.\nonumber\\
\end{eqnarray}
These are again set of inhomogeneous, nonlinear differential equations. Therefore considering
the r.h.s. as source, we may express solutions in terms of Green's functions as usual. From our earlier 
solutions (\ref{solgauge}) and (\ref{zeroth}) for gauge fields, it is quite evident
that only the last equation corresponding to $ \mathcal{A}^{(1)(\kappa^{(1)})}_{z} $ will
have a non trivial solution at the leading order in $ \kappa $. In the following we enumerate the
solutions at the leading order as,
\begin{eqnarray}
\mathcal{A}_{t}^{(1)(\kappa^{(1)})}&=&0\nonumber\\
\mathcal{A}_{i}^{(1)(\kappa^{(1)})}&=&0\nonumber\\
\mathcal{A}_{z}^{(1)(\kappa^{(1)})}&=& -(2r_0^{2})\int_{0}^{1}du^{''}\frac{\mathcal{J}(u^{''},
\textbf{x}'')}{u''}\times \int d\textbf{x}^{''}\mathcal{G}_s(u,u^{''};\textbf{x},\textbf{x}'')
\end{eqnarray}
where $ \mathcal{J}(u,\textbf{x}) $ is the source for $ \mathcal{A}_{z}^{(1)(\kappa^{(1)})} $ and is given by,
\begin{equation}
\mathcal{J}(u,\textbf{x})= \rho_{0}^{2(\kappa^{(0)})}\sigma ({\bf{x}})\mathcal{A}_{z}^{(0)(\kappa^{(1)})}+\frac{8u^{2}}{r_0^{2}}(\partial_{u}\mathcal{A}^{(0)(\kappa^{(0)})}_{t}\mathcal{F}^{(1)(\kappa^{(0)})}_{xy}+H_{c2} \partial_{u}\mathcal{A}^{(1)(\kappa^{(0)})}_{t}).
\end{equation}
Finally we write down the first order Maxwell equations (\ref{l}) at
the quadratic order in $ \kappa $ which are given as follows.

\textbf{Equations in the second order in $ \kappa $:}
\begin{eqnarray}
L_t\mathcal{A}_{t}^{(1)(\kappa^{(2)})}-\frac{16 u^{2}f(u)}{r_0^{2}}(\partial_{u}\mathcal{A}^{(0)(\kappa^{(1)})}_{z}\mathcal{F}^{(1)
(\kappa^{(0)})}_{xy}+H_{c2} \partial_{u}\mathcal{A}^{(1)(\kappa^{(1)})}_{z})\nonumber\\
=\frac{2r_0^{2}\sigma ({\bf{x}})}{u}(2\mathcal{A}_{t}^{(0)(\kappa^{(0)})}\rho_{0}^{(\kappa^{(0)})}\rho_{0}^{(\kappa^{(2)})}+\mathcal{A}_{t}^{(0)(\kappa^{(2)})}\rho_{0}^{2(\kappa^{(0)})})\nonumber\\
L_s\mathcal{A}^{(1)(\kappa^{(2)})}_{x}-16 u(\partial_{u}\mathcal{A}^{(0)(\kappa^{(0)})}_{t}\partial_{y}\mathcal{A}^{(1)(\kappa^{(1)})}_{z}-\partial_u \mathcal{A}^{(0)(\kappa^{(1)})}_{z}\partial_{y}\mathcal{A}^{(1)(\kappa^{(0)})}_{t})\nonumber\\
=\frac{2r_0^{2}\rho_{0}^{(\kappa^{(0)})}\rho_{0}^{(\kappa^{(2)})}}{u}\epsilon_{x}^{y}\partial_{y}\sigma(\textbf{x})\nonumber\\
L_s\mathcal{A}^{(1)(\kappa^{(2)})}_{y}-16 u(\partial_u \mathcal{A}^{(0)(\kappa^{(1)})}_{z}\partial_{x}\mathcal{A}^{(1)(\kappa^{(0)})}_{t}-\partial_{u}\mathcal{A}^{(0)(\kappa^{(0)})}_{t}\partial_{x}\mathcal{A}^{(1)(\kappa^{(1)})}_{z})\nonumber\\
= \frac{2r_0^{2}\rho_{0}^{(\kappa^{(0)})}\rho_{0}^{(\kappa^{(2)})}}{u}\epsilon_{y}^{x}\partial_{x}\sigma(\textbf{x})\nonumber\\
L_s\mathcal{A}^{(1)(\kappa^{(2)})}_{z}-16 u(\partial_{u}\mathcal{A}^{(0)(\kappa^{(0)})}_{t}\mathcal{F}^{(1)(\kappa^{(1)})}_{xy}+H_{c2} \partial_{u}\mathcal{A}^{(1)(\kappa^{(1)})}_{t})\nonumber\\
=\frac{2r_0^{2}\rho_{0}^{2(\kappa^{(0)})}\sigma ({\bf{x}})}{u}\mathcal{A}_{z}^{(0)(\kappa^{(2)})}.
\end{eqnarray}
Using our standard technique, we can again express these solutions in
terms of Green's function, which take the following form
\begin{eqnarray}
\mathcal{A}^{(1)(\kappa^{(2)})}_{i}=-(2r_0^{2})\int_{0}^{1}du'\int d\textbf{x}'
\frac{\mathcal{S}_{i}(u',\textbf{x}')}{u'}\mathcal{G}_{s}(u,u^{'};\textbf{x},\textbf{x}')
\end{eqnarray}
where $ \mathcal{S}_{i}(u,\textbf{x}) $ is the source for $ \mathcal{A}^{(1)(\kappa^{(2)})}_{i} $ and
is given by,
\begin{eqnarray}
\mathcal{S}_i= \epsilon_{i}^{j}\partial_j\chi (u,\textbf{x}).
\end{eqnarray}
It is really remarkable to note at this stage that even in the
presence of anomaly we can express the source as a total derivative of
some function $ \chi (u,\textbf{x}) $ which is given by,
\begin{equation}
\chi (u,\textbf{x})=\rho_{0}^{(\kappa^{(0)})}\rho_{0}^{(\kappa^{(2)})}\sigma(\textbf{x})+
\frac{8u^{2}}{r_0^{2}}\mathcal{H}(u,\textbf{x})
\end{equation}
where, $ \mathcal{H} $ is some function that depends on the gauge
fields and its derivatives
\begin{equation}
\mathcal{H}=\partial_{u}\mathcal{A}^{(0)(\kappa^{(0)})}_{t}
\mathcal{A}^{(1)(\kappa^{(1)})}_{z}-\partial_u 
\mathcal{A}^{(0)(\kappa^{(1)})}_{z}\mathcal{A}^{(1)(\kappa^{(0)})}_{t}.
\end{equation}
It is to be noted that we have not written out the solutions for $
\mathcal{A}_{t}^{(1)(\kappa^{(2)})} $ and $
\mathcal{A}_{z}^{(1)(\kappa^{(2)})} $ explicitly. The reason for this
is that we are interested to express our current upto order $
\kappa^{2} $ and for that we do not need this part of solution
according to our formula (\ref{j}).  Substituting all the above
solutions into (\ref{j}) finally we obtain\footnote{We have used the
  fact that $ \epsilon_{ij} $ is anti symmetric, $ r_0=1 $ and $
  \frac{\partial \mathcal{G}}{\partial x'} = -\frac{\partial
    \mathcal{G}}{\partial x}$.},
\begin{eqnarray}
  J_i&=&-\frac{\varepsilon}{16\pi G}\epsilon_i^{j}\partial_j 
\Xi +\frac{\varepsilon}{6\pi G} \epsilon_{i}^{j}\partial_{j} 
\Theta \equiv\epsilon_{i}^{j}\partial_{j}\Gamma ~~~(i=x,y)\nonumber\\
  J_z&=&-2\frac{\kappa \varepsilon}{4\pi G}\Sigma \label{j1}
\end{eqnarray}
This is the exact expression for the \textit{anomalous} current in the
presence of vortex lattice upto the quadratic order in $ \kappa $,
where we have combined both the entities $ \Xi $ and $ \Theta $ into a
single entity $ \Gamma $. Note that the quantities $ \Gamma $ and $
\Sigma $ are explicit functions of gauge fields and their
derivatives. In the following we give the full (analytic) expression
for these quantities
\begin{eqnarray}
  \Xi &=& \int d\textbf{x}'\sigma(\textbf{x}')\partial_u\int_{0}^{1}du'
  \frac{\rho_{0}^{2(\kappa^{(0)})}(u^{'})}{u'}\mathcal{G}_{s}(u,u^{'};\textbf{x},\textbf{x}')
  +2\kappa^{2}\int d\textbf{x}'\partial_u \int_{0}^{1} \frac{du'}{u'}
  \chi(u',\textbf{x}')\mathcal{G}_{s}(u,u^{'};\textbf{x},\textbf{x}')|_{u=0}\nonumber\\
  \Theta &=&
  \kappa^{2}(\mathcal{A}^{(0)(\kappa^{(0)})}_{t}\mathcal{A}^{(1)(\kappa^{(1)})}_{z}
  -\mathcal{A}^{(0)(\kappa^{(1)})}_{z}\mathcal{A}^{(1)(\kappa^{(0)})}_{t})|_{u=0}\nonumber\\ 
  \Sigma &=& \int d\textbf{x}' \partial_u \int_{0}^{1}du^{'}\frac{\mathcal{J}(u^{'},
    \textbf{x}')}{u'}\mathcal{G}_{s}(u,u^{'};\textbf{x},\textbf{x}')|_{u=0}\label{sigma}
\end{eqnarray}
Note that the quantity $ \Xi $ has two parts in it. The term that is
proportional to $ \kappa^{2} $ is arising purely due to the presence
of CS term. Whereas on the other hand, the first term is always there
even if we do not have any anomaly in our theory. Also one can note
that the quantity $ \Theta $ is arising solely due to the presence of
the CS term. Therefore this term will be present in the current only
if the anomaly is there.
 
From the above expressions (\ref{sigma}) it is quite evident that the
current thus obtained is non-local. On the other hand, in the usual GL
theory the current is expressed as a local function of the order
parameter. Therefore we are still few steps away from the standard GL
theory. However, one may arrive at the local expression for the
current in a certain limit which we discuss below.

From (\ref{sigma}) the reader might notice that there are non
localities associated with both the radial ($ u $) direction as well
as with the ($ x,y $) plane. In order to make these expressions local,
as a first step we extract the $ u $ dependence of the Green's
functions. We define a complete set of orthonormal basis such that,
\begin{eqnarray}
\mathcal{G}_{t}(u,u';\textbf{x},\textbf{x}')&=& \sum\limits_{\zeta}\xi_{\zeta}(u)
\xi_{\zeta}^{\dagger}(u')\tilde{\mathcal{G}}_{t}(\textbf{x},\zeta)\nonumber\\
\mathcal{G}_{s}(u,u';\textbf{x},\textbf{x}')&=& \sum\limits_{\lambda}\eta_{\lambda}(u)
\eta_{\lambda}^{\dagger}(u')\tilde{\mathcal{G}}_{s}(\textbf{x},\lambda)
\end{eqnarray}
with,
\begin{eqnarray}\label{lambdadf}
  \mathcal{L}_{t}\xi_{\zeta}(u)=\zeta
  \xi_{\zeta}(u);~~~\sum\limits_{\zeta}\xi_{\zeta}(u)\xi_{\zeta}^{\dagger}(u')=\delta(u-u');~~~\langle\xi_{\zeta}|\xi_{\zeta'}\rangle 
  = \delta_{\zeta\zeta'}\nonumber\\ 
  \mathcal{L}_{s}\eta_{\lambda}(u)=\lambda \eta_{\lambda}(u);~~~\sum\limits_{\lambda}\eta_{\lambda}(u)\eta_{\lambda}^{\dagger}(u')=\delta(u-u');~~~\langle\eta_{\lambda}|\eta_{\lambda'}\rangle = \delta_{\lambda\lambda'}\label{ls1}.
\end{eqnarray}
Here $ \tilde{\mathcal{G}}_{t}(\textbf{x},\lambda) $ and $ \tilde{\mathcal{G}}_{s}(\textbf{x},\lambda) $
are the Green's functions defined in the ($ x,y $) plane that basically satisfy the 
 equation of motion of the form,
\begin{eqnarray}
(\Delta - \wp^{2})\tilde{\mathcal{G}}(\textbf{x},\wp^{2}) &=& -\delta (\textbf{x})\label{green}
\end{eqnarray}
for any real positive value of $ \wp^{2} $.

The solution of (\ref{green}) may be expressed in terms of modified Bessel function
\begin{eqnarray}
\tilde{\mathcal{G}}(\textbf{x},\wp^{2})=\frac{1}{2\pi}K_0 (\wp |x|)
\end{eqnarray}
which satisfies the condition $ lim_{|\textbf{x}|\rightarrow \infty}|\tilde{\mathcal{G}}(\textbf{x})|<\infty $.

The eigen functions $\lbrace \xi_{\zeta}\rbrace $ and $
\lbrace\eta_{\lambda}\rbrace $ that form a complete orthonormal basis,
satisfy the following boundary conditions\footnote{Note that these
  boundary conditions are nothing but the artifact of our previously
  defined boundary conditions in (\ref{bc1}).}
\begin{eqnarray}
\xi_{\zeta}(u)|_{u=1}=\xi_{\zeta}(u)|_{u=0}=0\nonumber\\
\eta_{\lambda}(u)|_{u=1}=\eta_{\lambda}(u)|_{u=0}=0\label{bc2}.
\end{eqnarray}

In order to remove non localities in the ($ x,y $) plane\footnote{We
  are interested in removing the non localities associated with the ($
  x,y $) plane because of the fact that the non trivial condensate ($
  \sigma(\textbf{x}) $) which forms below a critical magnetic field ($
  H_{c2} $) is basically spanned in the ($ x,y $) plane. } we consider
the \textit{long wave length approximation} which comes with the
following idea: The Green's function varies at a length scale ($\sim
1/\sqrt{\lambda} $) small compared to that of the vortex solution
whose size could be determined by the length scale ($\xi $)fixed by the vortex
lattice. Therefore, in this long wave length approximation, we may
take our condensate to be uniform over the length scale on which the
Green's function fluctuates\footnote{According to the BCS theory, the
  usual superconductors are associated with small length scales
  characterized by the BCS coherence length and the mean free path. In
  presence of these small length scales the condensate does not remain
  uniform over the region where the gauge field fluctuates and as a
  result the current takes the non local form.}. Mathematically we may
state this long wave length approximation as,
  \begin{eqnarray}
 \frac{1}{\sqrt{\lambda}}\ll\xi.
  \end{eqnarray}
  This is indeed a
vital assumption which finally removes the non localities associated
with \textbf{x} and helps us to express our anomalous current locally
as a function of the order parameter ($ \sigma(\textbf{x}) $).

In this long wavelength approximation, we define the convolution ($ *
$) of the Green's function in the (x,y) plane as,
\begin{eqnarray}\label{longwlapprox}
[\tilde{\mathcal{G}}_{s}(\lambda)*\sigma](\textbf{x})=\int d\textbf{x}'
\tilde{\mathcal{G}}_{s}(\textbf{x}-\textbf{x}',\lambda)\sigma(\textbf{x}')=\frac{\sigma(\textbf{x})}{\lambda}.
\end{eqnarray}
where, $ \tilde{\mathcal{G}}_{s} $ is the Green's function satisfying,
\begin{eqnarray}
\int d\textbf{x}\tilde{\mathcal{G}}_{s}(\textbf{x})=\frac{1}{\lambda}.
\end{eqnarray}
Therefore, using the above approximations and definitions we finally
write down the current as a local function of the condensate $
\sigma(\textbf{x}) $. We give our results below.
\begin{eqnarray}
\Xi &=& \tilde{\tilde{\mathcal{N}}}_{\lambda}\sigma(\textbf{x})+2\kappa^{2}\mathcal{N}^{(0)(\kappa^{(2)})}_{\lambda(2)}\nonumber\\
\Theta &=&\mathcal{M}_{1}\sigma(\textbf{x})+\mathcal{M}_2 + \mathcal{M}_{3}\Delta \sigma(\textbf{x})+\mathcal{M}_{4}\sigma(\textbf{x})+\mathcal{M}_{5}\sigma(\textbf{x})\nonumber\\
\Sigma &=& \mathcal{P}_{1}\sigma(\textbf{x})+\mathcal{P}_2 + \mathcal{P}_{3}\Delta \sigma(\textbf{x})+\mathcal{P}_{4}\sigma(\textbf{x})
\end{eqnarray}
where $\mathcal{N}$, $ \mathcal{M} $ and $ \mathcal{P} $ are all constants which can be
expressed as summation over $ \lambda $ whose details are given in the Appendix. In
the following we shall concentrate only on $ \Xi $ and $ \Theta $
as we are interested in finding the current $ J_i $.

\subsection{Large $ \lambda $ Limit} \label{largelambda}

In this section we further simplify the expressions for the constants
$\mathcal{N}$ and $ \mathcal{M} $ taking into account the large
$\lambda$ limit which essentially means that we shall replace $
\lambda $ by $ \lambda_{min} $ in the sum and will be concerned only
with those terms that are mostly dominant in the series expansion of $
\lambda $. Considering this fact we finally obtain,
\begin{eqnarray}
\Xi &=&  \frac{\eta'_{\lambda_{min}}(0)}{\lambda_{min}}\left( \mathcal{F}(\lambda_{min},\zeta_{min})\sigma(\textbf{x})+\kappa^{2}\frac{H_{c2}\mathcal{C}}{\lambda_{min}}\right)\nonumber\\
\Theta &=& \frac{\eta_{\lambda_{min}}(0)}{\lambda_{min}}\left[\left
(\sum\limits_{i=7,8} \mathcal{C}_{i}+\frac{\mathcal{C}_{9}}
{\lambda_{min}}\Delta + \frac{H_{c2}}{\zeta_{min}}\mathcal{C}_{10}\right)\sigma 
(\textbf{x})+H_{c2}\mathcal{C}_{11}  \right]\label{theta}.
\end{eqnarray}
Where the function $ \mathcal{F}(\lambda_{min},\zeta_{min})\sigma(\textbf{x}) $ takes the following form
\begin{eqnarray}
\mathcal{F}(\lambda_{min},\zeta_{min})= \mathcal{C}_1+\kappa^{2}\left( \mathcal{C}_2+\frac{\mathcal{C}_3}
{\lambda_{min}}+\frac{\mathcal{C}_4}{\zeta_{min}}+\frac{\mathcal{C}_5}{\lambda_{min}^{2}}\Delta 
+\frac{H_{c2}\mathcal{C}_6}{\lambda_{min}\zeta_{min}}\right)\label{F}.
\end{eqnarray}
Here $ \mathcal{C}_{i} $s are all constants that could be estimated by knowing the behavior of
the eigen functions $\lbrace \xi_{\zeta}\rbrace $ and $ \lbrace\eta_{\lambda}\rbrace $ as well as
the radial function ($ \rho_{0}(u) $) near the horizon and the boundary of the AdS.

Our next task is to further simplify the expressions given in (\ref{theta}). From our earlier 
discussions in section 2 one can note that the vortex
solution could be expressed as $ \sigma (\textbf{x})=|\psi_0|^{2} $ where
$ \psi_0 $ is given by (\ref{psi0}). Using this relation, we 
finally express $ \Delta \sigma (\textbf{x}) $ in terms of the lowest energy 
 solution ($ \psi_0 $) of the GL equation, that essentially turns out to be,
\begin{eqnarray}
\Delta\sigma(\textbf{x})&=&(\partial_x^{2}+\partial_y^{2})\left( e^{-\frac{x^{2}}{\xi^{2}}}\vartheta_3^{\dagger}\vartheta_3\right)\nonumber\\
&=&(-2H_{c2} +4A_y^{2})|\psi_0|^2 -\frac{4ix}{a_y\xi^{3}}e^{-\frac{x^{2}}{\xi^{2}}}
\bigg(\frac{\partial}{\partial v^{*}}-\frac{\partial}{\partial v}\bigg)e^{\frac{x^{2}}{ \xi^{2}}}|\psi_0|^2
+\frac{4}{a_y^{2}\xi^{2}}e^{-\frac{x^{2}}{\xi^{2}}}\frac{\partial^2}{\partial v^{*}\partial v}
e^{\frac{x^{2}}{ \xi^{2}}}|\psi_0|^2\label{dsigma}\nonumber\\
\end{eqnarray}
where we have used the fact that  
\begin{eqnarray}
A_y= H_{c2}x=\frac{x}{\xi^{2}}.
\end{eqnarray}
Also from the
boundary conditions (\ref{bc2}) it is quite evident 
that all the terms those are proportional to $ \eta_{\lambda_{min}}(0) $
will ultimately vanish. In this sense the term $ \Theta $ does not 
exist at all and therefore we are solely left with $ \Xi $
that finally determines the GL current in the presence of anomaly.
 Using all these facts one can finally express $ \Xi $ as,
\begin{eqnarray}
\Xi =  {\cal A}_1 |\psi_0|^2 + {\cal A}_2 |\psi_0|^2 +\mathcal{A}_3 \label{xi}
\end{eqnarray}
 where the coefficients $ {\cal A}_1 $, $ {\cal A}_2 $ and $ {\cal A}_3 $ are given below\footnote{Note that
 the coefficients $ {\cal A}_2 $ and $ {\cal A}_3 $ arise solely due to the presence of the anomaly.},
\begin{eqnarray}\label{coefficients}
{\cal A}_1&=& \frac{\eta'_{\lambda_{min}}(0)}{\lambda_{min}}\left[\mathcal{C}_1+ \kappa^2 
\left(\mathcal{C}_2 + \frac{\mathcal{C}_3}
{\lambda_{min}}+\frac{\mathcal{C}_4}{\zeta_{min}}
+\frac{H_{c2}\mathcal{C}_6}{\lambda_{min}\zeta_{min}}\right)\right]\nonumber\\
\mathcal{A}_2 &=& \frac{\kappa^{2}\eta'_{\lambda_{min}}(0)\mathcal{C}_5}{\lambda_{min}^{3}}\left[(- 2 H_{c2}+ 4 A_{y}^2)+\bigg(
 -\frac{4ix}{a_y\xi^{3}}e^{-\frac{x^{2}}{\xi^{2}}}
\bigg(\frac{\partial}{\partial v^{*}}-\frac{\partial}{\partial v}\bigg)
+\frac{4}{a_y^{2}\xi^{2}}e^{-\frac{x^{2}}{\xi^{2}}}\frac{\partial^2}{\partial v^{*}\partial v}\bigg)
e^{\frac{x^{2}}{ \xi^{2}}} \right]  \nonumber\\
\mathcal{A}_3 &=&\frac{\kappa^{2}}{\lambda_{min}^{2}}\eta'_{\lambda_{min}}(0)H_{c2}\mathcal{C}.\label{A} 
\end{eqnarray}
 From the above expressions (\ref{dsigma}), (\ref{xi}) and (\ref{A}) and
  using the AdS/CFT prescription, it is now quite trivial 
  to compute the boundary current (\ref{j1}) 
 (which is basically the GL current for the present case) in the long wave length approximation, which
  takes the following form\footnote{Here we have re-scaled the current $ J_i $ by the factor $\frac{-\varepsilon}{16 \pi G}$.},
\begin{eqnarray}\label{Fcurrent}
J_i = {\cal A}_1\epsilon_i^j \pa_j \sigma(\textbf{x})
+\frac{\kappa^{2}\eta'_{\lambda_{min}}(0)\mathcal{C}_5}{\lambda_{min}^{3}} \epsilon_i^j \pa_j (\Delta \sigma(\textbf{x})).\label{jfinal}
\end{eqnarray} 
This is the final expression for the GL current in the presence of the
 global $ U(1) $ anomaly. From the above expression (\ref{jfinal}) it can
 be easily noticed that in the presence of the anomaly
 the usual GL current receives a non trivial
correction that arises at the quadratic order in ($ 1/\lambda $) with respect to 
the leading term\footnote{The anomaly corrections that appear in $ {\cal A}_1 $ is trivial since it is 
associated with the vortex solution $ \sigma(\textbf{x}) $.}. By nontrivial correction we
explicitly mean the term associated with derivative of
 the local function $ \Delta \sigma(\textbf{x}) $.
  The coefficient for this
non trivial correction term can be estimated by knowing
the constant  $ \mathcal{C}_{5} $, which is
of the following form \footnote{See the corresponding term $ \mathcal{N}^{(2)(\kappa^{(2)})}_{\lambda(2)} $
in the Appendix.},
\begin{eqnarray}
\mathcal{C}_{5}=
-128
\int_{0}^{1} du'u'\partial_{u'}\mathcal{A}^{(0)
(\kappa^{(0)})}_{t}(u')\eta_{\lambda_{min}'}(u')\eta_{\lambda_{min}}^{\dagger}(u')\nonumber\\
\times\int_{0}^{1} du''u''\partial_{u''}\mathcal{A}^{(0)(\kappa^{(0)})}_{t}(u'')
\eta_{\lambda_{min}''}(u'')\eta_{\lambda_{min}'}^{\dagger}(u'')\int_{0}^{1}
\frac{ du'''}{u'''}\rho_{0}^{2(\kappa^{(2)})}(u''')\eta_{\lambda_{min}''}^{\dagger}(u''')\label{c5}.
\end{eqnarray}
Before we proceed further, there are some important issues that has to be discussed 
at this stage. First of all considering the boundary behavior (\ref{bcpsi}) of
the scalar field $ \psi $ (which basically tells us about the
nature of the radial solution $ \rho_0(u) $ as $ u\rightarrow 0 $)we note
that all the integrals (over the radial coordinate $ u $)
that appear in (\ref{c5}) are finite, and therefore the coefficient
$ \mathcal{C}_{5} $ is a finite number.

Secondly and most importantly, looking at (\ref{ls}) and (\ref{ls1})
one can easily figure out that $ \sqrt{\lambda}\sim T $ where $ T $ is the
temperature of our system. Thus the non trivial term on
the r.h.s of the above equation (\ref{jfinal}) essentially corresponds
to a finite temperature correction to the usual GL current. This is the correction
to the current (due to the presence of the anomaly)
that goes as $(\sim1/T^{4}) $ w.r.t the leading term
and thereby highly suppressed at large temperatures. Thus following the
AdS/CFT prescription, for real
life superconductors one should realize similar effects 
at finite but non zero temperatures. This further suggests that the usual
GL current would be modified at finite non zero temperatures.

\subsection{Modifying GL theory}

Inspired by the entire analysis done so far, we are now in a position
to make the following comments regarding the usual GL theory for real
life superconductors: According to the AdS/CFT duality the usual GL
current ($ J_{GL} $) would receive a highly nontrivial correction once
we incorporate the effect of anomaly in the theory and the effect of
such correction could be measured only at finite (low)
temperatures. As the temperature of the sample is increased we would
get back to the usual GL current with some trivial correction due to
anomaly, due to the presence of $\mathcal{C}_2$ in ${\cal A}_1$ as in
evident from equations (\ref{coefficients}) and
(\ref{Fcurrent}). Therefore, based on this observation we propose the
following modification to the usual GL current for ordinary
superconductors,
\begin{eqnarray}
J_{GL}\rightarrow \epsilon_{i}^{j}\partial_j \sigma
(\textbf{x})+\kappa^{2}\epsilon_{i}^{j}\partial_j
(\Delta\sigma(\textbf{x}))\label{glmod} 
\end{eqnarray}
where the first term stands for the usual GL current
that is already familiar to us, while the
 second term on the r.h.s. of (\ref{glmod}) stands for the
non trivial correction that appears at finite but low temperatures.

\section{Free energy}
\label{sec:freeenergy}

Computation of free energy is always important in order to describe
the thermodynamic stability of a given system. A particular
configuration is stable if it possesses the minimum free energy. The
phase transition that we describe through out this paper, is basically
a \textit{second} order transition between a normal and a
superconducting phase in the presence of an external magnetic field.

In the usual GL theory, thermodynamically most favorable configuration
corresponds to a triangular lattice solution that minimizes the free
energy.  Since our boundary theory possesses a global $ U(1) $ anomaly
which arises due to the presence of the CS term in the bulk action,
therefore, it is naturally expected that in the present case, the free
energy would receive some non trivial corrections due to the presence
of this CS term. From the computations of this section we find that
this is indeed the case, where considering the large $ \lambda $
approximation we give the precise expression of the free energy up to
the quadratic order in $ \kappa $.

The free energy of a particular system is defined as the
\textit{onshell} action evaluated with appropriate counter terms,
\begin{equation}
F = - S_{os}\label{f1}.
\end{equation}
Let us first evaluate the (\textit{onshell}) action ($ S_{os} $)
(\ref{action}) for the scalar field. Note that the equation of motion
for the scalar field mat be written as,
\begin{eqnarray}
D_{a}^{2}\psi -m^{2}\psi = 0
\end{eqnarray}
where,
\begin{eqnarray}
D_a = \nabla_a -iA_a.
\end{eqnarray}
Using this we find that the action (\ref{action}) evaluated \textit{on shell} turns out to be,
\begin{equation}
S_{\psi}|_{os}=-\frac{1}{2}\int_{\partial M}d\Sigma_{a}\sqrt{-g}(\nabla^{a}-iA^{a})|\psi|^{2}.
\end{equation}
Here $ d\Sigma_{a} $ is the volume of the hyper surface $ \partial M $ which is the boundary of the AdS space ($ M $). In the following we consider various cases regarding the choice of $ \partial M $. 
\begin{itemize}
\item 
If we take $ a=t $, then $ \partial M $ will correspond to (past and future) space like surfaces where according to our choice of \textit{stationary} field configurations the time derivative of the fields vanishes.
 \item
 If we choose $ a=u $, then this integral vanishes at the horizon (as $ g^{uu}=0 $) and also at the boundary ($ u\rightarrow 0 $) due to the rapid fall off of the scalar field.
\item
 For the case $ a=i $, we choose $ x=constant $  hyper surfaces in such a way so that the condensate is highly suppressed due to the presence of the exponentially decaying factor\footnote{Note that, since $ \psi(x,y) \approx e^{i k_y y}f_0(x) $, therefore $ |\psi|^{2} $ will only depend on $ x $ (see (\ref{psinot})).}. 
\end{itemize}
Thus, considering all these facts the action for the scalar field
 evaluated \textit{onshell} turns out to be,
\begin{equation}
S_{\psi}|_{os}=0.
\end{equation}
Therefore we are finally left with the \textit{onshell} action for the gauge fields only,
\begin{eqnarray}
S_{os}= \int d^{5}x\sqrt{-g}\left(- \frac{1}{4}F^{2}+\frac{\kappa}{3}\frac{\epsilon^{abcde}}{\sqrt{-g}}A_aF_{bc}F_{de}\right)\label{s}.
\end{eqnarray}
In order to evaluate (\ref{s}), we expand it perturbatively in $ \varepsilon $
\begin{eqnarray}
S_{os}= S^{(0)}_{os}+\varepsilon S^{(1)}_{os}+\varepsilon^{2} S^{(2)}_{os}+\mathcal{O}(\varepsilon^{3}).\label{s1}
\end{eqnarray}
Where various terms on the r.h.s. of (\ref{s}) corresponds to
perturbations of the action at different order in $ \varepsilon $.
Our aim is to evaluate various terms in (\ref{s1}) using the zeroth as well as the first order Maxwell equations,
\begin{eqnarray}
\nabla_{b}F^{(0)ab}- \frac{\kappa}{\sqrt{-g}}\epsilon^{abcde}F^{(0)}_{bc}F^{(0)}_{de}&=&0\nonumber\\
\nabla_{b}F^{(1)ab}-2\frac{\kappa}{\sqrt{-g}} \epsilon^{abcde}F^{(0)}_{bc}F^{(1)}_{de}&=& j^{(1)a}\label{e1}
\end{eqnarray}
along with the orthogonality condition
\begin{eqnarray}
\int d^{5}x \sqrt{-g}j^{(1)a}A^{(1)}_{a}=0\label{e2}.
\end{eqnarray}
First of all, we may drop $ S_{os}^{(0)} $ because we are interested in computing the free energy corresponding to a non zero value of the condensate.

Next, using (\ref{e1}) and (\ref{e2}), we compute the first order correction to the action which turns out to be,
\begin{eqnarray}
S_{os}^{(1)} = -\int_{\partial M} d\Sigma_{u}\sqrt{-g}F^{(0)ub}A^{(1)}_{b}|_{u=0}=0
\end{eqnarray}
where we have used the fact that the sources corresponding to $ A_t $ and $ A_z $ are kept fixed at the boundary. As a result both $ A^{(i)}_{t} $ as well as $ A^{(i)}_{z} $ (i=1,2,..) vanishes as we approach the boundary of the AdS ($ u\rightarrow0 $). 

Finally we compute the second order correction to the action ($ S^{(2)}_{os} $), which turns out to be,
\begin{eqnarray}
S^{(2)}_{os}=-\int_{\partial M}d\Sigma_{u}\sqrt{-g}F^{(0)ub}A^{(2)}_{b}|_{u=0}-\frac{1}{2}\int_{\partial M}d\Sigma_{u}\sqrt{-g}F^{(1)ub}A^{(1)}_{b}|_{u=0}\nonumber\\
-2\kappa \int_{\partial M}d\Sigma_{u}\epsilon^{ubcde}A_{b}^{(0)}F^{(0)}_{cd}A^{(2)}_{e}|_{u=0}-2\kappa \int_{\partial M}d\Sigma_{u}\epsilon^{ubcde}A_{b}^{(0)}F^{(1)}_{cd}A^{(1)}_{e}|_{u=0}\label{e3}.
\end{eqnarray}
Following our previous arguments, one can note that the first and the third term on the r.h.s of (\ref{e3}) vanishes identically. On the other hand, using (\ref{j}) the second and the fourth term may be combined as,
\begin{eqnarray}
S^{(2)}_{os}=\frac{1}{2\varepsilon}\int_{\partial M}d\Sigma_{u} J^{i}A^{(1)}_{i}|_{u=0}.
\end{eqnarray}
Finally using (\ref{j1}), we arrive at the following expression for the \textit{onshell} action\footnote{Note that $ \Theta $ is zero at the boundary $ u=0 $.},
\begin{eqnarray}
S_{os}=\frac{\varepsilon^{2}}{2}\int_{\partial M}d\Sigma_{u} \epsilon_{ij}\partial_{j}\Xi A^{(1)}_{i}|_{u=0}=-\frac{\varepsilon^{2}}{2}\int_{\partial M}d\Sigma_{u}\Xi \epsilon_{ij}\partial_{j} a_{i}|_{u=0}=\frac{H_{c2}\varepsilon^{2}}{2}\int_{\Re^{2}}d\textbf{x} \Xi(\textbf{x})\label{sos}.
\end{eqnarray}
For any function $ f(\textbf{x}) $, let us define the following quantity
\begin{eqnarray}
\overline{f}=\frac{1}{V}\int_{\Re^{2}}d\textbf{x} f(\textbf{x})\label{xibar}
\end{eqnarray}
which represents the average of $ f(\textbf{x}) $ over the volume $ (V) $ in the $ (x,y) $
plane.

Substituting (\ref{sos}) into (\ref{f1}) and using the above definition (\ref{xibar}), the free energy($ F $)
per unit volume ($ V $) in the ($ x,y $) plane turns out to be,
\begin{equation}
F/V= - \frac{H_{c2}\varepsilon^{2}}{2}\overline{\Xi}\label{F1}
\end{equation}
where,
\begin{eqnarray}
\overline{\Xi} = {\cal A}_1 \overline{\sigma(\textbf{x})}
+\frac{\kappa^{2}\eta'_{\lambda_{min}}(0)\mathcal{C}_5}{\lambda_{min}^{3}}\overline{\Delta \sigma(\textbf{x})}\label{F2}.
\end{eqnarray}
This is the final expression for the GL free energy in the presence of a global $ U(1) $ anomaly.
From the above expressions (\ref{F1}) and (\ref{F2}) it should be clear by now that in the presence of a global
$ U(1) $ anomaly, the quantity $ F/V $
would be modified by a nontrivial factor. This effect is quite similar to that what we have found in the previous section while computing the GL current in the presence of $ U(1) $ anomaly. Like in the previous case we note that this nontrivial effect appears $ \mathcal{O}( \frac{1}{T^{4}}) $ w.r.t. the leading term in the expression and therefore would modify the original GL free energy only at finite (low) temperatures.

\section{Conclusions}\label{conclusion}

We shall conclude our discussion with few remarks.

The holographic computation of supercurrent turns out to be a
non-local quantity. However, in the large $\lambda$ limit, what we
considered in section \ref{largelambda}, the expression takes the form
of a local quantity. The large $\lambda$ correction essentially tells
that
\be
\frac{\sqrt{B_{c2}}}{T} \ll 1 .
\ee
Therefore the effect of CS term in the supercurrent is suppressed at
high temperature with respect to the leading term. 

One important fact about the final expression of super-current
(equation (\ref{glmod})) is that, only the CS term does not produce
$1/\lambda^2$ correction in the current. There are other terms which
also give $1/\lambda^2$ corrections. The source for these terms come
from large $\lambda$ correction of $\sigma(x)$:
\ben
\sigma({\bf x'}) \sim \sigma({\bf x})+ \frac{(\#)}{\sqrt{\lambda}}
\sigma'({\bf x}) + \cdots  
\een
When we substitutes the expansion of $\sigma(x')$ in equation
(\ref{longwlapprox}) we get correction of order $1/\lambda$,
$1/\lambda^2$ and so on. However, our goal in this paper is not to
find large $\lambda$ correction to the super-current but to compute
the effect of CS term on holographic super-current. Therefore, we did
not consider these terms in our final result.

{\bf {Acknowledgements :}}

We would like to thank Suhas Gangadharaiah and Krishnendu Sengupta for
valuable discussions. SD acknowledges the hospitality of NIKHEF,
Amsterdam, where part of the work was done. NB would like to thank the
hospitality of IISER Bhopal at the final stage of this work. Research
of NB is supported by DST Ramanujan Grant. Finally we are thankful to
the people of India for their support to Science.

\vspace{1cm}

\appendix

\noindent
{\bf \large Appendix}

\section{Calculation of $\Xi$, $\Theta$ and $\Sigma$}

The appendix consists of all important formulea that apprear whole computing the boundary current. 
The expressions are long important and hence we provide them here. In equation (\ref{j1}), the two quantities that 
appear in the $i-th$ component of the current $J_i$ are $\Xi$ and $\Theta$ and the one 
that apprears in $J_z$ is $\Sigma$ . Below, we outline the steps 
to calculate these three quantities.\\
\textbf{Calculation of $\Xi  $ :}
\begin{eqnarray}
\Xi = \Xi^{(\kappa^{(0)})}+2 \kappa^{2}\Xi^{(\kappa^{(2)})}
\end{eqnarray}
\begin{eqnarray}
\Xi^{(\kappa^{(0)})}= \mathcal{N}^{(\kappa^{(0)})}_{\lambda} \sigma(\textbf{x})
\end{eqnarray}
where,
\begin{eqnarray}
\mathcal{N}^{(\kappa^{(0)})}_{\lambda}= \sum\limits_{\lambda}\frac{\eta'_{\lambda}(0)}
{\lambda}\int_{0}^{1}du'\frac{\rho_{0}^{2(\kappa^{(0)})}(u')}{u'}\eta_{\lambda}^{\dagger}(u')
\end{eqnarray}
\textbf{Note:}
\begin{eqnarray}
\Xi^{(\kappa^{(2)})}=\Xi_{(1)}^{(\kappa^{(2)})}+\Xi_{(2)}^{(\kappa^{(2)})}
\end{eqnarray}
\begin{eqnarray}
\Xi_{(1)}^{(\kappa^{(2)})}=\mathcal{N}_{\lambda(1)}^{(\kappa^{(2)})}\sigma(\textbf{x})
\end{eqnarray}
\begin{eqnarray}
\mathcal{N}_{\lambda(1)}^{(\kappa^{(2)})}=\sum\limits_{\lambda}\frac{\eta'_{\lambda}(0)}{\lambda}\int_{0}^{1}\frac{du'}{u'^{2}}\rho_{0}^{(\kappa^{(0)})}(u')\rho_{0}^{(\kappa^{(2)})}(u')\eta_{\lambda}^{\dagger}(u')
\end{eqnarray}
Let us calculate the following
\begin{eqnarray}
-16\int d\textbf{x}'\partial_u \int d\textbf{x}^{''} \int_{0}^{1} du'u'\partial_{u'}\mathcal{A}^{(0)
(\kappa^{(0)})}_{t}\int_{0}^{1}du^{''}
\frac{\mathcal{J}(u^{''},\textbf{x}'')}{u''}\times \mathcal{G}_s(u',u^{''};\textbf{x}',\textbf{x}'')
\mathcal{G}_s(u,u^{'};\textbf{x},\textbf{x}')\nonumber\\
\end{eqnarray}
It has three terms.

\textbf{First term:}
\begin{eqnarray}
\mathcal{N}_{\lambda(2)}^{(1)(\kappa^{(2)})}\sigma(\textbf{x})
\end{eqnarray}
where,
\begin{eqnarray}
\mathcal{N}_{\lambda(2)}^{(1)(\kappa^{(2)})}=-16\sum\limits_{\lambda,\lambda'}
\frac{\eta_{\lambda}'(0)}{\lambda}\int_{0}^{1} du'u'\partial_{u'}\mathcal{A}^{(0)(\kappa^{(0)})}_{t}(u')
\eta_{\lambda}^{\dagger}(u')\nonumber\\
\times \eta_{\lambda'}(u') \int_{0}^{1}\frac{du''}{u''}\rho_{0}^{2(\kappa^{(0)})}(u'')
\mathcal{A}^{(0)(\kappa^{(1)})}_{z}(u'')
\frac{\eta_{\lambda}'(u')}{\lambda'}\eta_{\lambda'}^{\dagger}(u'')
\end{eqnarray}

Also we note that,
\begin{eqnarray}
\mathcal{F}^{(1)(\kappa^{(0)})}_{xy}(u,\textbf{x}) = -H_{c2} + r_0^{2}
\sum\limits_{\lambda}\frac{\eta_{\lambda}(u)}{\lambda}\int_{0}^{1} du'
\frac{\rho_{0}^{2(\kappa^{(2)})}}{u'}\eta_{\lambda}^{\dagger}(u')\Delta \sigma(\textbf{x})
\end{eqnarray}

\textbf{Second term:}
\begin{eqnarray}
\frac{128H_{c2}}{r_0^{2}}\sum\limits_{\lambda,\lambda'}\frac{\eta_{\lambda}'(0)}{\lambda}
\int_{0}^{1} du'u'\partial_{u'}\mathcal{A}^{(0)(\kappa^{(0)})}_{t}(u')
\frac{\eta_{\lambda'}(u')}{\lambda'}\eta_{\lambda}^{\dagger}(u')\int_{0}^{1}
du''u''\partial_{u''}\mathcal{A}^{(0)(\kappa^{(0)})}_{t}(u'')\eta_{\lambda'}^{\dagger}(u'')\nonumber\\
+\mathcal{N}^{(2)(\kappa^{(2)})}_{\lambda(2)}\Delta\sigma(\textbf{x})\nonumber\\
\end{eqnarray}

where,
\begin{eqnarray}
\mathcal{N}^{(2)(\kappa^{(2)})}_{\lambda(2)}=-128\sum\limits_{\lambda,\lambda',\lambda''}
\frac{\eta_{\lambda}'(0)}{\lambda}\int_{0}^{1} du'u'\partial_{u'}\mathcal{A}^{(0)
(\kappa^{(0)})}_{t}(u')\frac{\eta_{\lambda'}(u')}{\lambda'}\eta_{\lambda}^{\dagger}(u')\nonumber\\
\times\int_{0}^{1} du''u''\partial_{u''}\mathcal{A}^{(0)(\kappa^{(0)})}_{t}(u'')
\frac{\eta_{\lambda''}(u'')}{\lambda''}\eta_{\lambda'}^{\dagger}(u'')\int_{0}^{1}
\frac{ du'''}{u'''}\rho_{0}^{2(\kappa^{(2)})}(u''')\eta_{\lambda''}^{\dagger}(u''')
\end{eqnarray}

\textbf{Third term:}
\begin{eqnarray}
\mathcal{N}^{(3)(\kappa^{(2)})}_{\lambda(2)}\sigma(\textbf{x})
\end{eqnarray}

where,
\begin{eqnarray}
\mathcal{N}^{(3)(\kappa^{(2)})}_{\lambda(2)}= 16^2 H_{c2}\sum\limits_{\lambda,\lambda',\zeta}
\frac{\eta_{\lambda}'(0)}{\lambda}\int_{0}^{1}du'u'\partial_{u'}\mathcal{A}^{(0)
(\kappa^{(0)})}_{t}(u')\frac{\eta_{\lambda'}(u')}{\lambda'}\eta_{\lambda}^{\dagger}(u')\nonumber\\
\times\int_{0}^{1}du''u''\frac{\xi'_{\zeta}(u'')}{\zeta}\eta_{\lambda'}^{\dagger}(u'')
\int_{0}^{1}\frac{du'''}{u'''}\rho_{0}^{2(\kappa^{(0)})}(u''')\mathcal{A}^{(0)
(\kappa^{(0)})}_{t}(u''')\xi^{\dagger}_{\zeta}(u''')
\end{eqnarray}

where,

\textbf{Last Term:}
\begin{eqnarray}
\mathcal{N}^{(4)(\kappa^{(2)})}_{\lambda(2)}\sigma(\textbf{x})
\end{eqnarray}
where,
\begin{eqnarray}
\mathcal{N}^{(4)(\kappa^{(2)})}_{\lambda(2)}=-16\sum\limits_{\lambda,\zeta}\frac{\eta'_{\lambda}(0)}
{\lambda}\int_{0}^{1}du'u'\partial_{u'}\mathcal{A}^{(0)(\kappa^{(1)})}_{z}(u')
\frac{\xi_{\zeta}}{\zeta}(u')\eta_{\lambda}^{\dagger}(u')\nonumber\\
\times\int_{0}^{1}du''\frac{\rho_{0}^{2(\kappa^{(0)})}}{u''}(u'')\mathcal{A}^{(0)
(\kappa^{(0)})}_{t}(u'')\xi^{\dagger}_{\zeta}(u'')
\end{eqnarray}

Finally adding all these terms we find,
\begin{eqnarray}
\Xi^{(\kappa^{((2)})}=\tilde{\mathcal{N}}^{(\kappa^{((2)})}_{\lambda}\sigma(\textbf{x})+\mathcal{N}^{(0)(\kappa^{(2)})}_{\lambda(2)}
\end{eqnarray}

where,
\begin{eqnarray}
\tilde{\mathcal{N}}^{(\kappa^{((2)})}_{\lambda}=\mathcal{N}^{(\kappa^{(2)})}_{\lambda(1)}+\mathcal{N}^{(1)(\kappa^{(2)})}_{\lambda(2)}+\mathcal{N}^{(2)(\kappa^{(2)})}_{\lambda(2)}\Delta +\mathcal{N}^{(3)(\kappa^{(2)})}_{\lambda(2)}+\mathcal{N}^{(4)(\kappa^{(2)})}_{\lambda(2)}
\end{eqnarray}
and,
\begin{eqnarray}
\mathcal{N}^{(0)(\kappa^{(2)})}_{\lambda(2)}=\frac{128 H_{c2}}{r_0^{2}}\sum\limits_{\lambda,\lambda'}
\frac{\eta_{\lambda}'(0)}{\lambda} \int_{0}^{1} du'u'\partial_{u'}\mathcal{A}^{(0)
(\kappa^{(0)})}_{t}(u')\frac{\eta_{\lambda'}(u')}{\lambda'}\eta_{\lambda}^{\dagger}(u')\int_{0}^{1}
du''u''\partial_{u''}\mathcal{A}^{(0)(\kappa^{(0)})}_{t}(u'')
\eta_{\lambda'}^{\dagger}(u'')\nonumber\\
\end{eqnarray}
\textbf{Final Result:}
\begin{eqnarray}
\Xi = \tilde{\tilde{\mathcal{N}}}_{\lambda}\sigma(\textbf{x})+2\kappa^{2}\mathcal{N}^{(0)(\kappa^{(2)})}_{\lambda(2)}
\end{eqnarray}
where,
\begin{eqnarray}
\tilde{\tilde{\mathcal{N}}}_{\lambda}=\mathcal{N}^{(\kappa^{(0)})}_{\lambda}+2\kappa^{2}\tilde{\mathcal{N}}^{(\kappa^{((2)})}_{\lambda}
\end{eqnarray}

\textbf{Calculation of $ \Theta $:}

\begin{eqnarray}
\mathcal{A}^{(0)(\kappa^{(0)})}_{t}|_{u=0}&=&q\\
\mathcal{A}^{(0)(\kappa^{(1)})}_{z}|_{u=0}&=& C
\end{eqnarray}
\textbf{First term:}
\begin{eqnarray}
\mathcal{M}_{1}\sigma(\textbf{x})
\end{eqnarray}
where,
\begin{eqnarray}
\mathcal{M}_{1} = -\frac{\kappa^{2}q C}{3 \pi G}\sum\limits_{\lambda}\frac{\eta_{\lambda}(0)}{\lambda}\int_{0}^{1}\frac{du'}{u'}\rho_{0}^{2(\kappa^{(0)})}(u')\eta_{\lambda}^{\dagger}(u')
\end{eqnarray}

\textbf{Second term:}

\begin{equation}
\mathcal{M}_2 + \mathcal{M}_{3}\Delta \sigma(\textbf{x})
\end{equation}

where,
\begin{eqnarray}
\mathcal{M}_{2}&=& \frac{8\kappa^{2} H_{c2}q}{3 \pi G}\sum\limits_{\lambda}\frac{\eta_{\lambda}(0)}{\lambda}\int_{0}^{1}du' u'\partial_{u'}\mathcal{A}^{(0)(\kappa^{(0)})}_{t}(u')\eta_{\lambda}^{\dagger}(u')\\
\mathcal{M}_{3}&=&-\frac{8\kappa^{2}q}{3 \pi G}\sum\limits_{\lambda,\lambda'}\frac{\eta_{\lambda}(0)}{\lambda}\int_{0}^{1}du' u'\partial_{u'}\mathcal{A}^{(0)(\kappa^{(0)})}_{t}(u')\eta_{\lambda}^{\dagger}(u')\frac{\eta_{\lambda'}(u')}{\lambda'} \int_{0}^{1}\frac{du''}{u''}\rho_{0}^{2(\kappa^{(2)})}(u'')\eta_{\lambda'}^{\dagger}(u'')\nonumber\\
\end{eqnarray}

\textbf{Third term:}
\begin{equation}
\mathcal{M}_{4}\sigma(\textbf{x})
\end{equation}
where,
\begin{eqnarray}
\mathcal{M}_{4}=\frac{16\kappa^{2} H_{c2}q}{3 \pi G}\sum\limits_{\lambda,\zeta}\frac{\eta_{\lambda}(0)}{\lambda}\int_{0}^{1}du' u'\frac{\xi'^{\dagger}_{\zeta}(u')}{\zeta}\eta_{\lambda}^{\dagger}(u')\int_{0}^{1}\frac{du''}{u''}\rho_{0}^{2(\kappa^{(0)})}(u'')\mathcal{A}^{(0)(\kappa^{(0)})}_{t}(u'')\xi^{\dagger}_{\zeta}(u'')\nonumber\\
\end{eqnarray}
\textbf{Last term:}
\begin{equation}
\mathcal{M}_{5}\sigma(\textbf{x})
\end{equation}
where,
\begin{eqnarray}
\mathcal{M}_{5}=-\frac{\kappa^{2}C}{3\pi G}\sum\limits_{\lambda}\frac{\eta_{\lambda}(0)}{\lambda}\int_{0}^{1}\frac{du'}{u'}\rho_{0}^{2(\kappa^{(0)})}(u')\mathcal{A}^{(0)(\kappa^{(0)})}_{t}(u')\eta_{\lambda}^{\dagger}(u')
\end{eqnarray}

\textbf{Calculation of $ \Sigma $:}

\textbf{First Term:}
\begin{equation}
\mathcal{P}_{1}\sigma(\textbf{x})
\end{equation}
where,
\begin{eqnarray}
\mathcal{P}_{1}=\sum\limits_{\lambda}\frac{\eta_{\lambda}'(0)}{\lambda}\int_{0}^{1}\frac{du'}{u'^{2}}\rho_{0}^{2(\kappa^{(0)})}(u^{'})\mathcal{A}^{(0)(\kappa^{(1)})}_{z}(u')\eta_{\lambda}^{\dagger}(u')
\end{eqnarray}

\textbf{Second Term:}
\begin{equation}
\mathcal{P}_{2}+\mathcal{P}_{3}\Delta\sigma(\textbf{x})
\end{equation}
where,
\begin{eqnarray}
\mathcal{P}_{2} &=&-4 H_{c2}\sum\limits_{\lambda}\frac{\eta_{\lambda}'(0)}{\lambda}\int_{0}^{1}\frac{du'}{u'^{4}}\partial_{u'}\mathcal{A}^{(0)(\kappa^{(0)})}_{t}(u')\eta_{\lambda}^{\dagger}(u')\nonumber\\
\mathcal{P}_{3}&=&4\sum\limits_{\lambda,\lambda'}\frac{\eta_{\lambda}'(0)}{\lambda}\int_{0}^{1}\frac{du'}{u'^{4}}\partial_{u'}\mathcal{A}^{(0)(\kappa^{(0)})}_{t}(u')\eta_{\lambda}^{\dagger}(u')\frac{\eta_{\lambda'}}{\lambda'}(u')\int_{0}^{1}\frac{du''}{u''^{2}}\rho_{0}^{2(\kappa^{(2)})}(u^{''})\eta_{\lambda'}^{\dagger}(u'')\nonumber\\
\end{eqnarray}

\textbf{Third Term:}
\begin{equation}
\mathcal{P}_{4}\sigma(\textbf{x})
\end{equation}
where,
\begin{eqnarray}
\mathcal{P}_{4}=-8\sum\limits_{\lambda,\lambda'}\frac{\eta_{\lambda}'(0)}{\lambda}\int_{0}^{1}\frac{du'}{u'^{4}}\frac{\eta_{\lambda'}'(u')}{\lambda'}\eta_{\lambda}^{\dagger}(u')\int_{0}^{1}\frac{du''}{u''^{2}}\rho_{0}^{2(\kappa^{(0)})}(u^{''})\mathcal{A}^{(0)(\kappa^{(0)})}_{t}(u'')\eta_{\lambda'}^{\dagger}(u'')\sigma(\textbf{x})\nonumber\\
\end{eqnarray}

Using the above expressions, we can get the explicit form of the boundary current.


\begin{thebibliography}{99}
\bibitem{Gubser:2005ih} 
  S.~S.~Gubser,
  ``Phase transitions near black hole horizons,''
  Class.\ Quant.\ Grav.\  {\bf 22}, 5121 (2005)
  [hep-th/0505189].
\bibitem{Gubser:2008px} 
  S.~S.~Gubser,
  ``Breaking an Abelian gauge symmetry near a black hole horizon,''
  Phys.\ Rev.\ D {\bf 78}, 065034 (2008)
  [arXiv:0801.2977 [hep-th]].
\bibitem{Hartnoll:2008vx} 
  S.~A.~Hartnoll, C.~P.~Herzog and G.~T.~Horowitz,
  ``Building a Holographic Superconductor,''
  Phys.\ Rev.\ Lett.\  {\bf 101}, 031601 (2008)
  [arXiv:0803.3295 [hep-th]].
\bibitem{Hartnoll:2008kx} 
  S.~A.~Hartnoll, C.~P.~Herzog and G.~T.~Horowitz,
  ``Holographic Superconductors,''
  JHEP {\bf 0812}, 015 (2008)
  [arXiv:0810.1563 [hep-th]].
\bibitem{Maeda:2009vf} 
  K.~Maeda, M.~Natsuume and T.~Okamura,
  ``Vortex lattice for a holographic superconductor,''
  Phys.\ Rev.\ D {\bf 81}, 026002 (2010)
  [arXiv:0910.4475 [hep-th]].
  \bibitem{Albash:2008eh} 
  T.~Albash and C.~V.~Johnson,
  ``A Holographic Superconductor in an External Magnetic Field,''
  JHEP {\bf 0809}, 121 (2008)
  [arXiv:0804.3466 [hep-th]].
  \bibitem{Albash:2009iq} 
  T.~Albash and C.~V.~Johnson,
  ``Vortex and Droplet Engineering in Holographic Superconductors,''
  Phys.\ Rev.\ D {\bf 80}, 126009 (2009)
  [arXiv:0906.1795 [hep-th]].
  \bibitem{Salvio:2012at} 
  A.~Salvio,
  ``Holographic Superfluids and Superconductors in Dilaton-Gravity,''
  JHEP {\bf 1209}, 134 (2012)
  [arXiv:1207.3800 [hep-th]].
  \bibitem{Domenech:2010nf} 
  O.~Domenech, M.~Montull, A.~Pomarol, A.~Salvio and P.~J.~Silva,
  ``Emergent Gauge Fields in Holographic Superconductors,''
  JHEP {\bf 1008}, 033 (2010)
  [arXiv:1005.1776 [hep-th]].
  \bibitem{Montull:2012fy} 
  M.~Montull, O.~Pujolas, A.~Salvio and P.~J.~Silva,
  ``Magnetic Response in the Holographic Insulator/Superconductor Transition,''
  JHEP {\bf 1204}, 135 (2012)
  [arXiv:1202.0006 [hep-th]].
\bibitem{Montull:2009fe} 
  M.~Montull, A.~Pomarol and P.~J.~Silva,
  ``The Holographic Superconductor Vortex,''
  Phys.\ Rev.\ Lett.\  {\bf 103}, 091601 (2009)
  [arXiv:0906.2396 [hep-th]].
\bibitem{Zayas:2011dw} 
  L.~A.~Pando Zayas and D.~Reichmann,
  Phys.\ Rev.\ D {\bf 85}, 106012 (2012)
  [arXiv:1108.4022 [hep-th]].
   \bibitem{Bertlmann:1996xk} R.~A.~Bertlmann,
  ``Anomalies in quantum field theory,''
  Oxford, UK: Clarendon (1996) 566 p. (International series of monographs on physics: 91)
\bibitem{Harvey:2005it} 
  J.~A.~Harvey,
  ``TASI 2003 lectures on anomalies,''
  hep-th/0509097.
\bibitem{Bastianelli:2006rx} 
  F.~Bastianelli and P.~van Nieuwenhuizen,
  ``Path integrals and anomalies in curved space,''
  Cambridge, UK: Univ. Pr. (2006) 379 P.
\bibitem{Bilal:2008qx} 
  A.~Bilal,
  ``Lectures on Anomalies,''
  arXiv:0802.0634 [hep-th].
  \bibitem{Gynther:2010ed}
 A.~Gynther, K.~Landsteiner, F.~Pena-Benitez and A.~Rebhan,
 ``Holographic Anomalous Conductivities and the Chiral Magnetic Effect,''
 JHEP {\bf 1102}, 110 (2011)
 [arXiv:1005.2587 [hep-th]].
  \bibitem{Banerjee:2012cr} 
  N.~Banerjee, S.~Dutta, S.~Jain, R.~Loganayagam and T.~Sharma,
  ``Constraints on Anomalous Fluid in Arbitrary Dimensions,''
  JHEP {\bf 1303}, 048 (2013)
  [arXiv:1206.6499 [hep-th]].
  
  \bibitem{Maldacena:1997re} 
  J.~M.~Maldacena,
  ``The Large N limit of superconformal field theories and supergravity,''
  Adv.\ Theor.\ Math.\ Phys.\  {\bf 2}, 231 (1998)
  [hep-th/9711200]. 
  \bibitem{Witten:1998qj} 
  E.~Witten,
  ``Anti-de Sitter space and holography,''
  Adv.\ Theor.\ Math.\ Phys.\  {\bf 2}, 253 (1998)
  [hep-th/9802150].
  \bibitem{GL} V. L. Ginzburg and L. D. Landau, Zh. Eksp. Teor. Fiz.
20, 1064 (1950).
\bibitem{gorkov}Gor’kov L P 1986 Zh. Eksp. Teor. Fiz. 90 1478 (1986 Sov.
Phys. JETP 63 866).
\bibitem{Bardeen:1957mv} 
  J.~Bardeen, L.~N.~Cooper and J.~R.~Schrieffer,
  ``Theory of superconductivity,''
  Phys.\ Rev.\  {\bf 108}, 1175 (1957).
\bibitem{tinkham}M. Tinkham. Dover. Introduction to Superconductivity, 2nd edition : New York (1996).
  
  
  
  
  
  
  








\end{thebibliography}
\end{document}